\documentclass[a4paper, pdftex,pra,aps,onecolumn,twoside,nofootinbib, accepted=2021-06-15]{quantumarticle}
\pdfoutput=1
\usepackage[numbers,sort&compress]{natbib}

\usepackage[T1]{fontenc}
\usepackage[utf8]{inputenc}
\usepackage[colorlinks=true,linkcolor=blue,citecolor=red,plainpages=false,pdfpagelabels]{hyperref}
\usepackage{float}
\usepackage{braket}
\usepackage{indentfirst}
\usepackage{enumerate}
\usepackage{url}
\usepackage[dvipsnames]{xcolor}
\usepackage[stable]{footmisc}
\usepackage{fancyhdr}
\usepackage[all]{xy}
\usepackage{graphicx}
\usepackage{youngtab}
\usepackage{mathtools}
\usepackage{enumerate}
\usepackage{textcomp}
\usepackage{amsmath,amsfonts,amssymb,amsthm}
\usepackage[mathscr]{eucal}
\usepackage{tcolorbox}
\theoremstyle{plain}
\newtheorem{theorem}{Theorem}
\newtheorem{lemma}[theorem]{Lemma}
\newtheorem{proposition}[theorem]{Proposition}

\newtheorem{definition}[theorem]{Definition}

\newtheoremstyle{note}{\topsep}{\topsep}{\slshape}{}{\scshape}{}{ }{}
\theoremstyle{note}

\newtheorem*{theorem*}{Theorem}

\newcommand\tr{\operatorname{Tr}}

\newcommand{\<}{\langle}
\renewcommand{\>}{\rangle}

\newcommand\be{\begin{equation}}
\newcommand\ee{\end{equation}}
\newcommand\bea{\begin{array}}
	\newcommand\eea{\end{array}}
\newcommand\ben{\begin{eqnarray}}
\newcommand\een{\end{eqnarray}}
\newcommand\ot{\otimes}

\newcommand\bei{\begin{itemize}}
	\newcommand\eei{\end{itemize}}
\newcommand\bee{\begin{enumerate}}
	\newcommand\eee{\end{enumerate}}

\begin{document} 	
	\title{ Optimal Multi-port-based Teleportation Schemes \vspace{0.4cm}}

\author{Marek Mozrzymas}
\affiliation{Institute for Theoretical Physics, University of Wrocław,
  		50-204~Wrocław, Poland}
\author{Micha{\l} Studzi\'nski}
\affiliation{Institute of Theoretical Physics and Astrophysics, National Quantum Information Centre,  University of Gda\'nsk,\\ \mbox{80-952~Gda\'nsk}, Poland}
\author{Piotr Kopszak}
\affiliation{Institute for Theoretical Physics, University of Wrocław,
  		50-204~Wrocław, Poland}

	\begin{abstract}
In this paper, we introduce optimal versions of a  multi-port based teleportation scheme allowing to send a large amount of quantum information. We fully characterise probabilistic and deterministic case by presenting expressions for the average probability of success and entanglement fidelity. In the probabilistic case, the final expression depends only on global parameters describing the problem, such as the number of ports $N$, the number of teleported systems $k$, and local dimension $d$. It allows us to show square improvement in the number of ports with respect to the non-optimal case. We also show that the number of teleported systems can grow when the number $N$ of ports increases as $o(N)$ still giving high efficiency. In the deterministic case, we connect entanglement fidelity with the maximal eigenvalue of a generalised teleportation matrix. In both cases the optimal set of measurements and the optimal state shared between sender and receiver is presented.  All the results are obtained by formulating and solving primal and dual SDP problems, which due to existing symmetries can be solved analytically. We use extensively tools from representation theory and formulate new results that could be of the separate interest for the potential readers.
\end{abstract}
 \maketitle			 
%
\section{Introduction}
Quantum teleportation introduced originally in~\cite{bennett_teleporting_1993} is one of the building blocks of many quantum processing and communication protocols, as well as it gives us a deeper understanding of quantum physics~\cite{boschi_experimental_1998,gottesman_demonstrating_1999,gross_novel_2007, jozsa_introduction_2005, pirandola_advances_2015, raussendorf_one-way_2001, zukowski_event-ready-detectors_1993}. It employs shared entanglement for transfer of an unknown quantum state between two spatially separated laboratories, classical communication between parties, and depending on it unitary correction operation on the receiver's side.  However, the requirement of correction is a limiting  factor and it was natural to ask whether it is possible to get rid of this last step in the teleportation protocol. The first protocol of such kind has been introduced in~\cite{ishizaka_asymptotic_2008,ishizaka_quantum_2009} and called the port-based teleportation (PBT). The lack of correction in the last step allows for applications where the ordinary teleportation fails. One of the most prominent consequences of PBT, besides the model of universal programmable quantum processor~\cite{ishizaka_asymptotic_2008,ishizaka_quantum_2009,Banchi2020} are results regarding non-local quantum computation and quantum cryptography with new cryptographic attack reducing the amount of consumed entanglement~\cite{beigi_konig}, establishing the connection between communication complexity and the Bell inequality violation~\cite{buhrman_quantum_2016} or fundamental limitations on quantum channels distinguishability~\cite{limit}.

In PBT protocol both parties share a \textit{resource state} composed of $N$ pairs of maximally entangled states, each called \textit{port} (see the left panel of Figure~\ref{fig1}). Alice, in order to send to Bob an unknown state of a particle applies a joint measurement on her halves of ports and a state to be teleported, getting classical outcome $1\leq i\leq N$, transmitted further by a classical channel to Bob. This classical message indicates the port on which the teleported state arrives and no further correction is needed. 

We can distinguish two variants of the protocol, in both cases the optimal resource state as well as the set of measurements employed by Alice is different: \textit{Deterministic protocol:} In this variant, an unknown system is always transmitted to receiver but the transmission is imperfect. To learn about the quality of the teleported systems we compute the fidelity between the input and the output systems. Here, we focus on the problem of how well the teleportation channel transmits quantum correlations, so we ask about the entanglement fidelity.
\textit{Probabilistic protocol:} In this variant, the transmission is always perfect but there is a non-zero probability of failure of the whole process. To learn the efficiency of the protocol we want to know the probability of success of the scheme. 

There exists also a different version of PBT, the so-called \textit{optimal version}, where Alice optimises jointly measurements and the resource state before she runs the protocol (see left panel of Figure~\ref{fig1}). This results in shared global state that is no longer maximally entangled and allows for higher efficiency in every variant.

However, in every version of PBT, the perfect transmission with no probability of failure (probabilistic PBT) or with unit fidelity (deterministic PBT) is possible only in the asymptotic limit $N\rightarrow \infty$, which is a consequence of the no-go theorem for universal processor with finite system size~\cite{Nielsen1997}. This fact requires an answer to a question \textit{How well we can transmit a state with finite amount or resources (shared entangled pairs)?}  For example in deterministic qubit scheme~\cite{ishizaka_quantum_2009} the entanglement fidelity $F$ scales as $1-1/N^2$ in optimal protocol and as $1-1/N$ in non-optimal one. In probabilistic qubit version~\cite{ishizaka_quantum_2009} the probability of success $p$ scales as $1-1/N$ in optimal protocol and as $1-1/\sqrt{N}$ in non-optimal scheme. In every variant we have square improvement when moving to optimal procedure. For the full analysis of asymptotic performance of PBT scheme in all variants, for an arbitrary dimension $d$,  we refer to~\cite{majenz}.

All the above results have been derived when one considers teleportation of a single quantum system. When parties want to transmit a composite system, composed of $k$ subsytems, they can apply different variants of PBT: ordinary PBT with dimension of the port large enough, ordinary PBT applied many times with a new resource state in every step, recycling scheme for PBT~\cite{strelchuk_generalized_2013}, where the same resource state is exploited multiple times, or packaged version of PBT~\cite{strelchuk_generalized_2013,Kopszak2020}, where $N$ ports are divided into $k$ parts and parties run $k$ ordinary PBT protocols with $N/k$ ports each. The comparison of discussed variants is presented in papers~\cite{strelchuk_generalized_2013,Kopszak2020}.

There is also another protocol, allowing for teleportation of multiparty state in one go called multi-port based teleportation (MPBT), see the right panel of Figure~\ref{fig1}. The very first idea of such protocol has been described briefly in~\cite{strelchuk_generalized_2013} and fully analysed its non-optimal version in recent papers~\cite{Kopszak2020,stud2020A}. In this variant of teleportation the parties  can transfer a composite system or a number of states in one go with efficiency higher than in the all variants of the ordinary PBT described above by paying a price in mild correction on receiver's side in a form of ports permutation (in particular please see~\cite{Kopszak2020} for comparison).
However, up to now, only non-optimal case of MPBT has been considered for deterministic, as well as for probabilistic case. In this paper, we fill this gap by presenting optimal version of MPBT protocol in deterministic and probabilistic variant, together with discussion about resources consumption. Our results require new non-trivial extensions of tools coming from representation theory, in particular results concerning the representation theory of the algebra of partially transposed permutation operators, successfully applied to characterisation of PBT protocol for higher dimensions~\cite{Studzinski2017,StuNJP,MozJPA}, as well as to non-optimal MPBT~\cite{stud2020A}. In the next section, we briefly summarise results, separating fully technical contribution  from those related to MPBT.

\section{Our contribution}
We present analysis of both optimal probabilistic and optimal deterministic multi-port-based teleportation schemes. In particular our results contain:
\begin{enumerate}[1)]
\item Rigorous definition of the optimal multi-port based teleportation schemes, in both probabilistic and deterministic case.
    \item In optimal probabilistic scheme we derive closed expression for the average probability of success of the teleportation process. The final expression depends \textit{only} on global parameters like the number of teleported systems $k$, total number of ports $N$ and underlying dimension $d$ of the Hilbert space. The optimal protocol allows for better performance than the non-optimal one, while consuming less amount of entanglement.  We discuss the possible functional dependence $k=k(N)$  between the number of teleported systems and the total number of ports, still allowing parties for teleportation with high probability. In particular, we prove that parties can achieve $p=1$, for $N\rightarrow \infty$, whenever $k(N)=o(N)$. This is square improvement comparing to the non-optimal case discussed in~\cite{Kopszak2020} and it is the best possible functional dependence.    In the end we give explicit form of the resource state shared by parties and measurements performed by Alice to run the optimal protocol.
    \item In optimal deterministic scheme we show that the entanglement fidelity describing performance of the protocol can be expressed by the maximal eigenvalue of the  \textit{generalised teleportation matrix}, which is non-trivial generalisation of teleportation matrix discussed in the standard optimal port-based teleportation~\cite{StuNJP}.  We discuss the basic properties of this object. We compare our result numerically with non-optimal determinisitc MPBT, getting improvement in performance. Additionally, we present explicit form of the optimal resource state shared between Alice and Bob and the optimal measurements  performed by Alice. In this case we perform better than the non-optimal multi-port based protocol introduced in~\cite{stud2020A}, consuming less entanglement in the resource state. 
    \item We prove new, additional facts emerging from representation theory of symmetric group, as well as, from properties of recently studied algebra of partially transposed permutation operators $\mathcal{A}_n^{(k)}(d)$~\cite{stud2020A}, where $n=N+k$.  In particular we prove: a) Lemma~\ref{coef} for computing overlaps and partial traces between irreducible projectors of the algebra $\mathcal{A}_n^{(k)}(d)$ with basis vectors of irreducible representations of $S(n-2k)$. b) Theorem~\ref{thm} establishing the relation between squared  multiplicities of irreducible representations of groups $S(n-k)$ and $S(n-2k)$. 
    
    Since these results could be of potential interest for readers we dedicate a separate section to present them. 
\end{enumerate}
In both cases, i.e. the probabilistic and deterministic protocol, in order to find the final expression for probability of success, as well as for entanglement fidelity, we have to formulate primal and dual problems for semidefinite programming (SDP). These formulations and solutions explore extensively properties of the algebra of partially transposed permutation operators $\mathcal{A}_n^{(k)}(d)$. Symmetries induced by the mentioned algebra allow us solve optimisation problems analytically and show that solutions coming from the primal and the dual problems coincide, providing the exact values.

\section{Optimal version of the protocol}
\label{protocol}
The multi-port-based teleportation protocol has been introduced initially in papers~\cite{strelchuk_generalized_2013,Kopszak2020,stud2020A}, here for the completeness of the paper and reader's convenience  we recall it and formulate its optimal version, keeping the original notation. Additionally, we formulate primal and dual problems allowing us to find optimal values of entanglement fidelity, probability of success, and sets of optimal measurements.

\begin{figure}[h]
\centering
\includegraphics[width=.45\linewidth]{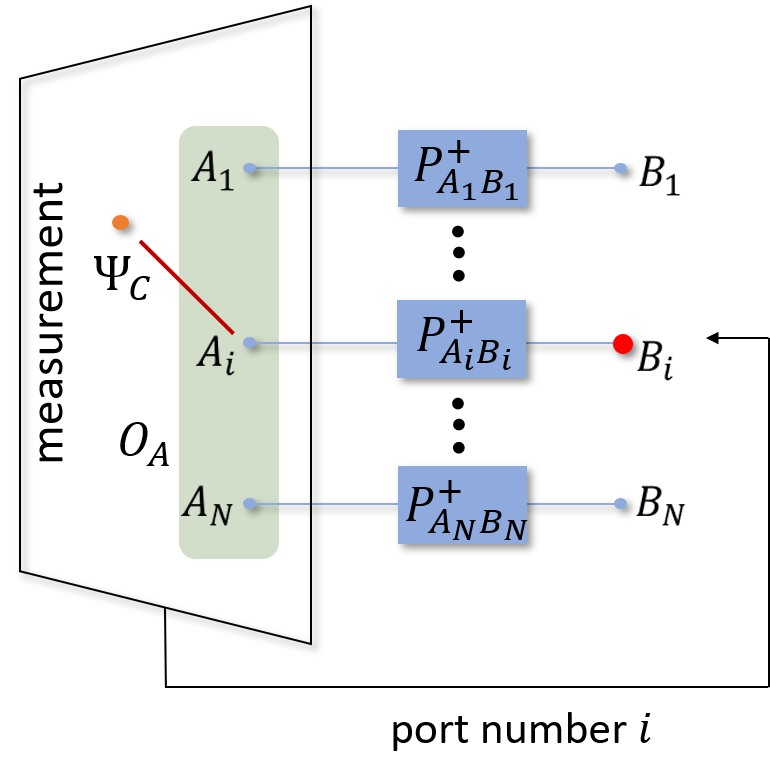} \ 
	\includegraphics[width=.45\linewidth]{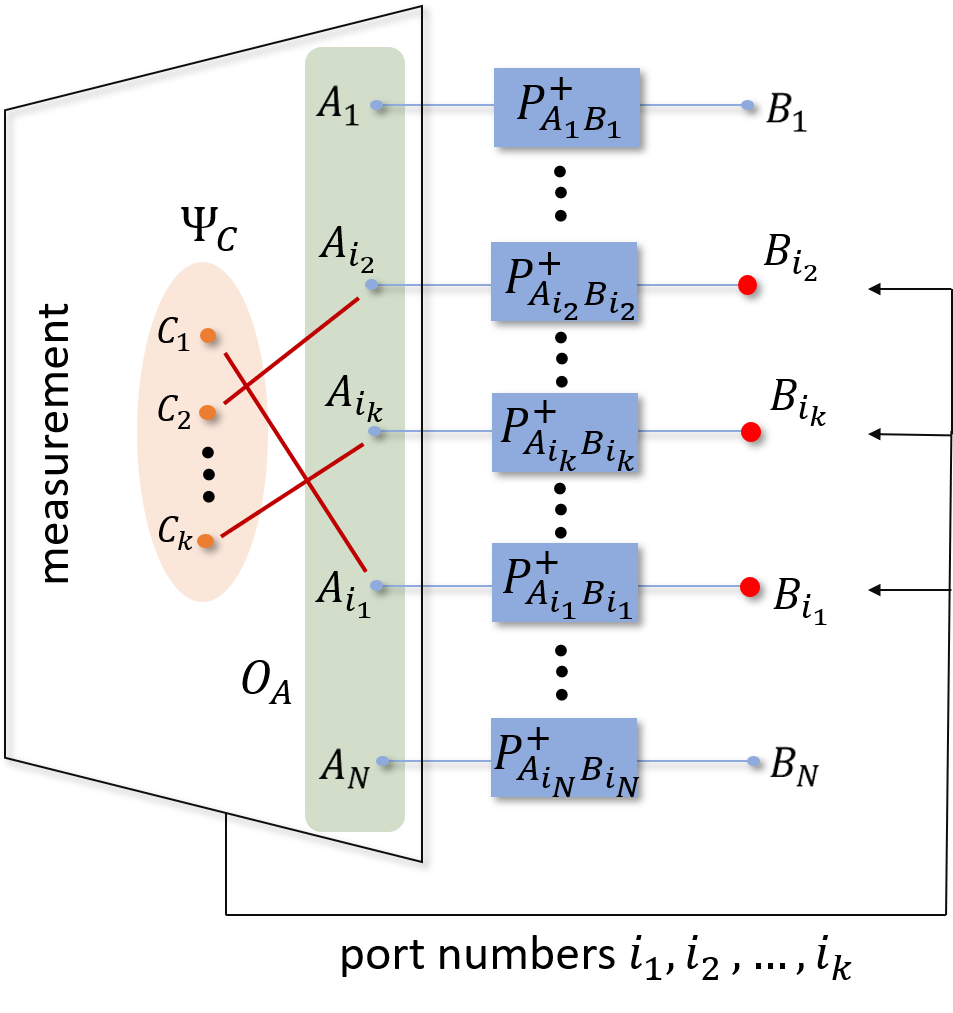}
	\caption{Figure presents configuration for the optimal port-based teleportation (left) and optimal  multi-port based teleportation scheme (right). By green rectangle, in both cases,  we denote the operation $O_A=O_{A_1A_2\cdots A_N}$ which Alice applies on her part of the resource state to increase capabilities of all variants of the teleportation protocol. In the standard PBT protocol, Alice applies a joint measurement on the system to be teleported and her halfs of maximally entangled pairs affected by $O_A$. She gets a classical outcome $1\leq i\leq N$ transmitted further by a classical channel to Bob and indicating on which port the teleported state has arrived.
	In the multi-port variants, after application of $O_A$ she performs a joint measurements on systems $C=C_1C_2\cdots C_k$ and $A=A_1A_2\cdots A_N$ getting an outcome $i_1,i_2,\ldots,i_k$. This outcome is sent to Bob and indicates on which ports the teleported state arrived (red dots). To recover the state Bob has to just permute the $k$ ports in the right order, indicated by the mentioned classical information. Please notice that operation $O_A$ is different in the standard PBT and its multi-port version.}
	\label{fig1}
\end{figure}

The task for two parties, Alice and Bob, is to transmit a composite unknown state $\Psi_C=\Psi_{C_1C_2\cdots C_k}$, where $k\geq 1$, or $k$ separate states $\psi_{C_1}\ot \psi_{C_2}\ot \cdots \ot \psi_{C_k}$ in one go. In the initial configuration the parties share $N$ maximally entangled pairs, called later the resource state. In contrast to previous versions of MPBT, now before a joint measurement applied on Alice's part of the resource state and the state for the teleportation, she is allowed to apply a global operation $O_A=O_{A_1A_2\cdots A_N}$ on her half of the resource state (see Figure~\ref{fig1}). After application of $O_A$ the resulting resource state is of the form
\be
\label{resource0}
|\Phi\>_{AB}=\left(O_A\ot \mathbf{1}_B \right) |\phi^+\>_{A_1B_1}\ot|\phi^+\>_{A_2B_2}\ot \cdots \ot |\phi^+\>_{A_NB_N},
\ee
where $\mathbf{1}_B$ is the identity operator acting on Bob's side and $|\phi^+\>_{A_jB_j}=(1/\sqrt{d})\sum_{i=1}^d|i\>_{A_j}|i\>_{B_j}$ denotes $d-$dimensional maximally entangled state between systems $A_j$ and $B_j$ for $1\leq j \leq N$. We set  $\tr \left( O_AO_A^{\dagger}\right) =d^N$ to obey normalisation of the resource state. Notice that the state $|\Phi\>_{AB}$ is no longer maximally entangled in the cut $A:B$, whenever $O_A\neq \mathbf{1}_A$ and $O_A$ is not an unitary operation. For notation simplicity we introduce the following set of $k-$tuples
\be
\mathcal{I}\equiv\left\lbrace \mathbf{i}=(i_1,i_2,\ldots,i_k) \ : \ \forall 1\leq l\leq k  \ i_l=1,\ldots,N  \ \text{and} \  i_1\neq i_2\neq \cdots \neq i_k \right\rbrace 
\ee
consists of all possible classical outcomes obtained by Alice during her measurements. The cardinality of $\mathcal{I}$ is $|\mathcal{I}|=k!\binom{N}{k}=\frac{N!}{(N-k)!}$, so it also corresponds to number of all measurements. 
Every Alice's measurement is described by positive operator valued measure (POVM) satisfying summation rule $\sum_{\mathbf{i}\in\mathcal{I}}\widetilde{\Pi}_{\mathbf{i}}^{AC}=\mathbf{1}_{AC}$.
Defining by $\Phi_{AB}=|\Phi\>\<\Phi|_{AB}$ and by $P^+_{A_jB_j}=|\phi^+\>\<\phi^+|_{A_jB_j}$, for $1\leq j \leq N$, we write that the teleportation channel $\mathcal{N}$ acts on the state $\Psi_C$ in the following manner:
\be
\label{ch1}
\begin{split}
	\mathcal{N}\left(\Psi_C \right)&=\sum_{\mathbf{i}\in \mathcal{I}}\tr_{A\bar{B}_{\mathbf{i}}C}\left[ \sqrt{\widetilde{\Pi}_{\mathbf{i}}^{AC}}\left(\Phi_{AB}\ot \Psi_C \right)\sqrt{\widetilde{\Pi}_{\mathbf{i}}^{AC}}^{\dagger}\right] \\
	&=\sum_{\mathbf{i}\in \mathcal{I}}\tr_{AC}\left[\widetilde{\Pi}_{\mathbf{i}}^{AC}\left(\left(O_A\ot \mathbf{1}_{B_{\mathbf{i}}} \right)\tr_{\bar{B}_{\mathbf{i}}}\bigotimes_{j=1}^N P^+_{A_jB_j}  \left(O_A^{\dagger}\ot \mathbf{1}_{B_{\mathbf{i}}} \right)\ot \Psi_C \right)\right]\\
	&=\sum_{\mathbf{i}\in \mathcal{I}} \tr_{AC}\left[\widetilde{\Pi}_{\mathbf{i}}^{AC} \left(\left(O_A\ot \mathbf{1}_{B_{\mathbf{i}}} \right)\sigma_{\mathbf{i}}^{AB} \left(O_A^{\dagger}\ot \mathbf{1}_{B_{\mathbf{i}}} \right) \ot \Psi_C\right)\right].
\end{split}
\ee
The notation $\tr_{\bar{B}_{\mathbf{i}}}$ denotes partial trace over all  subsystems except those on positions $\mathbf{i}=(i_1,i_2,\ldots,i_k)$.
 The partial trace $\tr_{\bar{B}_{\mathbf{i}}}$ in~\eqref{ch1} can be evaluated and it is equal to
\be
\label{sigma}
\sigma^{AB}_{\mathbf{i}}\equiv \tr_{\bar{B}_{\mathbf{i}}}\left(P^+_{A_1B_1}\ot P^+_{A_2B_2}\ot \cdots \ot P^+_{A_NB_N} \right)_{B_{\mathbf{i}}\rightarrow \widetilde{B}}=\frac{1}{d^{N-k}}\mathbf{1}_{\bar{A}_{\mathbf{i}}}\ot P^+_{A_{\mathbf{i}}\widetilde{B}},
\ee
with $B_{\mathbf{i}}\rightarrow \widetilde{B}=B_{N+1}\cdots B_{N+k}$ meaning that we assign $\widetilde{B}$ to the teleported state, whichever the $\mathbf{i}$. Further, we  refer to these states as signals. Every operator
$P^+_{A_{\mathbf{i}}\widetilde{B}}$ is then a tensor product of projectors on maximally entangled states between respective subsystems:
\begin{equation}
P^+_{A_{\mathbf{i}}\widetilde{B}}=P^+_{A_{i_1}B_{N+1}}\ot P^+_{A_{i_2}B_{N+2}}\ot \cdots \ot P^+_{A_{i_k}B_{N+k}}.
\end{equation}

In the deterministic scheme, to learn about its efficiency we can ask how well quantum correlations are transmitted through. More precisely, we have to compute entanglement fidelity between the input and the output sate, when Alice teleports subsystem $C$ from maximally entangled state $P^+_{DC}\cong P_{D_1C_1}^+\ot P_{D_2C_2}^+\ot \cdots \ot P_{D_kC_k}^+$ to Bob and keeps subsystem $D$ with her. The entangled fidelity after application of the channel $\mathcal{N}$ reads
\be
\begin{split}
F&=\tr\left[P^+_{DB}\left(\mathbf{1}_D\ot \mathcal{N}_C\right)P^+_{DC}\right]=\frac{1}{d^{2k}}\sum_{\mathbf{i}\in\mathcal{I}}\tr\left[\left(O_A^{\dagger}\ot \mathbf{1}_{B_{\mathbf{i}}} \right)\widetilde{\Pi}_{\mathbf{i}}^{AC} \left(O_A\ot \mathbf{1}_{B_{\mathbf{i}}} \right)\sigma_{\mathbf{i}}^{AB} \right]\\
&=\frac{1}{d^{2k}}\sum_{\mathbf{i}\in\mathcal{I}}\tr\left[\Pi_{\mathbf{i}}^{AC}\sigma_{\mathbf{i}}^{AB} \right],
\end{split}
\ee
where $P^+_{DB}\cong P_{D_1B_1}^+\ot P_{D_2B_2}^+\ot \cdots \ot P_{D_kB_k}^+$, and $\Pi_{\mathbf{i}}^{AC}=\left(O_A^{\dagger}\ot \mathbf{1}_{B_{\mathbf{i}}} \right)\widetilde{\Pi}_{\mathbf{i}}^{AC} \left(O_A\ot \mathbf{1}_{B_{\mathbf{i}}} \right)$ are rotated versions of the measurements.

In the probabilistic scheme, to learn about its efficiency we have to compute the average probability of success. In this case the teleportation channel  is the same as in~\eqref{ch1}, but it is trace non-preserving, since Alice has access to an additional POVM corresponding to failure. The probability of success in the multi-port case then reads
\be
p_{succ}=\frac{1}{d^{N+k}}\sum_{\mathbf{i}\in\mathcal{I}}\tr\left[\left(O_A^{\dagger}\ot \mathbf{1}_C \right)\widetilde{\Pi}_{\mathbf{i}}^{AC} \left(O_A\ot \mathbf{1}_C \right)\right]=\frac{1}{d^{N+k}}\sum_{\mathbf{i}\in\mathcal{I}}\tr\left[\Pi_{\mathbf{i}}^{AC}\right].
\ee
Please take into account that in the both cases the set of (rotated) measurements, as well as, the operation $O_A$ is different. We do not distinguish them in notation since it is always clear from the context in which scenario we work in. Our goal is to determine the explicit form of rotated measurements and operation $O_A$, and evaluate expressions for the entanglement fidelity and averaged probability of success.

To do so, in the next section we extensively exploit symmetries of the protocol, in particular operators describing it like signal states and measurements.  Let us focus here a little bit more on the symmetries exhibit by the protocol under investigation~\footnote{Most of the information in this part of the section are taken from~\cite{stud2020A}, we incorporate them here for the reader's convenience.}. First let us begin with the following simple observation. Namely, any bipartite maximally entangled state $P^+_{XY}$, between systems $X$ and $Y$ can be viewed as a swap operator $V[(X,Y)]$ exchanging systems $X$ and $Y$ distorted by partial transposition $t_Y$ with respect to system $Y$:
\begin{equation}
\label{PtoV'}
 P^+_{XY}=\frac{1}{d}V^{t_Y}[(X,Y)]=\frac{1}{d}V^{t_Y}_{XY}. 
\end{equation}
Moreover, every such maximally entangled state is $U\ot \overline{U}$ invariant, where the bar denotes complex conjugation of an element $U$ of the unitary group $\mathcal{U}(d)$.
Having the above, from the set of all signals $\{\sigma_{\mathbf{i}}^{AB}\}_{\mathbf{i}\in\mathcal{I}}$ we distinguish one corresponding to index $\mathbf{i}_0=(N-2k+1,N-2k+2,\ldots,N-k)$. Taking into account that the total number of systems is $n=N+k$ the signal state $\sigma^{AB}_{\mathbf{i}_{0}}$ corresponding to the index $\mathbf{i}_0$ is of the following form
\be
\label{canonical}
\begin{split}
	\sigma^{AB}_{\mathbf{i}_{0}}=\frac{1}{d^{n-k}}\mathbf{1}_{\bar{A}_{\mathbf{i}_0}}\ot P^+_{A_{n-2k+1},B_n}\ot P^+_{A_{n-2k+2},B_{n-1}}\ot \cdots \ot P^+_{A_{n-k},B_{n-k+1}}.
\end{split}
\ee
Using relation~\eqref{PtoV'} we rewrite~\eqref{canonical} in a more friendly form for the further analysis:
\be
\label{canonical2}
	\sigma^{AB}_{\mathbf{i}_{0}}=\frac{1}{d^{n}}\mathbf{1}_{\bar{A}_{\mathbf{i}_0}}\ot V^{t_{B_n}}_{A_{n-2k+1},B_n}\ot V^{t_{B_{n-1}}}_{A_{n-2k+2},B_{n-1}}\ot \cdots \ot V^{t_{B_{n-k+1}}}_{A_{n-k},B_{n-k+1}}:=\frac{1}{d^{n}}V^{(k)},
\ee
where
\be
\begin{split}
	&V^{(k)}:= \mathbf{1}_{\bar{A}_{\mathbf{i}_0}}\ot V^{t_{B_n}}_{A_{n-2k+1},B_n}\ot V^{t_{B_{n-1}}}_{A_{n-2k+2},B_{n-1}}\ot \cdots \ot V^{t_{B_{n-k+1}}}_{A_{n-k},B_{n-k+1}},\\
	&(k):= t_n \circ t_{n-1}\circ \cdots \circ t_{n-k+1},
\end{split}
\ee
and $\circ$ denotes composition of maps.
Now notice that any other signal state corresponding to some index $\mathbf{i}\in \mathcal{I}$ can be obtained from $\sigma^{AB}_{\mathbf{i}_{0}}$ by applying a some permutation belonging to the coset $\mathcal{S}_{n,k}\equiv \frac{S(n-k)}{S(n-2k)}$, i.e. signals are covariant with respect to this set or even complete permutation group $S(n-k)$. The same of course holds also when we start from different index than $\mathbf{i}_0$ and obtain the same conclusion, but this choice is the most natural and very useful further in this section. In general situation we have then:
\begin{equation}
\label{cov_sig}
V(\pi)\sigma^{AB}_{\mathbf{i}}V^{\dagger}(\pi)=\sigma^{AB}_{\pi(\mathbf{i})},\qquad \forall \ \pi\in S(n-k),
\end{equation}
where permutation operators $V(\pi)$, for $\pi \in S(N)$, permute factors in the space $\mathcal{H\equiv (\mathbb{C}}^{d})^{\otimes n}$ according to the following rule:
\be
\forall \pi \in S(n)\qquad V(\pi )|e_{1}\>\otimes |e_{2}\>\otimes
\cdots \otimes |e_{n}\>=|e_{\pi ^{-1}(1)}\>\otimes |e_{\pi
^{-1}(2)}\>\otimes \cdots \otimes |e_{\pi ^{-1}(n)}\>,
\ee
where  $\{|e_{i}\>\}_{i=1}^{d}$ is an orthonormal basis of the space $\mathcal{\mathbb{C}}^{d}.$
Except the covariance in~\eqref{cov_sig} all signals also satisfy relations:
\begin{equation}
\label{sig_inv}
\begin{split}
&\left[U^{\ot (n-k)}\ot \overline{U}^{\ot k}, \sigma^{AB}_{\mathbf{i}}\right]=0,\qquad \forall \ U\in\mathcal{U}(d),\\
&\left[V(\pi),\sigma^{AB}_{\mathbf{i}}\right]=0,\qquad \forall \ \pi\in S(n-2k).
\end{split}
\end{equation}
The same relations, i.e. covaraince~\eqref{cov_sig} and invariance~\eqref{sig_inv} hold also for the measurements in non-optimal deterministic and probabilistic MPBT scheme discussed in~\cite{stud2020A}. In this manuscript we show that the same holds for measurements in optimised versions of MPBT protocol.

Using the definition of the signal state $\sigma_{\mathbf{i}_0}^{AB}$ given in~\eqref{canonical2} through the operator $V^{(k)}$, together with covariance property of the signals in~\eqref{cov_sig}, one defines the MPBT operator $\rho= \sum_{\mathbf{i}\in \mathcal{I}}\sigma_{\mathbf{i}}^{AB}$ encoding properties of deterministic MPBT protocol (see Appendix~\ref{AppB} for the details):
\be
\label{PBT1}
\rho=\sum_{\mathbf{i}\in \mathcal{I}}\sigma_{\mathbf{i}}^{AB}=\frac{1}{d^n}\sum_{\tau \in \mathcal{S}_{n,k}}V(\tau^{-1})V^{(k)}V(\tau),
\ee
where the sum runs over all permutations $\tau$ from the coset $\mathcal{S}_{n,k}\equiv \frac{S(n-k)}{S(n-2k)}$. The operator $\rho$ is invariant with respect to action of any permutation from $S(n-k)$ as well as with respect to action of elements from $U^{\ot (n-k)}\ot \overline{U}^{\ot k}$.

The discussed symmetries imply also that the whole teleportation channel $\mathcal{N}$ defined through expression~\eqref{ch1} is invariant with respect to action of $U^{\ot (n-k)}\ot \overline{U}^{\ot k}$. Indeed, denoting by $\omega_{DB}=(\mathbf{1}_D\ot \mathcal{N}_C)P_{DC}^+$ a state isomorphic to channel, we write the following chain of equalities:
\begin{equation}
\begin{split}
&\left(U^{\ot (n-k)}\ot \overline{U}^{\ot k}\right)\omega_{DB}\left(U^{\ot (n-k)}\ot \overline{U}^{\ot k}\right)^{\dagger}\\
&=\sum_{\mathbf{i}\in \mathcal{I}} \tr_{AC}\left[\left(U^{\ot (n-k)}\ot \overline{U}^{\ot k}\right)\left(O_A\ot \mathbf{1}_{B_{\mathbf{i}}} \right)\widetilde{\Pi}_{\mathbf{i}}^{AC}\left(O_A^{\dagger}\ot \mathbf{1}_{B_{\mathbf{i}}} \right) \left(\sigma_{\mathbf{i}}^{AB}  \ot P_{DC}^+\right)\left(U^{\ot (n-k)}\ot \overline{U}^{\ot k}\right)^{\dagger}\right]\\
&=\sum_{\mathbf{i}\in \mathcal{I}} \tr_{AC}\left[\left(U^{\ot (n-k)}\ot \overline{U}^{\ot k}\right)\left(O_A\ot \mathbf{1}_{B_{\mathbf{i}}} \right)\widetilde{\Pi}_{\mathbf{i}}^{AC}\left(O_A^{\dagger}\ot \mathbf{1}_{B_{\mathbf{i}}} \right)\left(U^{\ot (n-k)}\ot \overline{U}^{\ot k}\right)^{\dagger} \left(\sigma_{\mathbf{i}}^{AB}  \ot P_{DC}^+\right)\right].
\end{split}
\end{equation}
In the non-optimal versions of MPBT discussed in~\cite{stud2020A}, i.e. when $O_A=\mathbf{1}_A$, the measurements $\widetilde{\Pi}_{\mathbf{i}}^{AC}$ commute with $U^{\ot (n-k)}\ot \overline{U}^{\ot k}$ showing that indeed the resulting channel is invariant with respect to $U^{\ot (n-k)}\ot \overline{U}^{\ot k}$. The same property holds for the channel performing the optimal teleportation (with non-trivial $O_A$). This is proven later in the text, where the symmetries of the protocol together with optimisation procedures require the optimal choice of the measurements to be invariant with respect to $U^{\ot (n-k)}\ot \overline{U}^{\ot k}$.

\section{Formulation of optimisation problems}
 Having  the formal description of the protocol we are in a position to formulate semidefinite problems (SDP) for deterministic and probabilistic variant of MPBT scheme.   We define both primal and dual problems, where solution of primal one upper bounds the real value, while solution of dual problem gives a lower bound on it. In both cases we are looking for any feasible solutions, so it is enough to assume that both operation $O_A$, measurements $\Pi_{\mathbf{i}}$, and any other auxiliary operators exhibit the same type of symmetries as signals $\sigma_{\mathbf{i}}$ and MPBT operator $\rho$. Later we show that in both variants of MPBT protocol, solutions of primal and dual problems coincide giving exact expressions for the real values of the entanglement fidelity and probability of success. This property is called as a strong duality property. Symmetries of the protocol discussed in Section~\ref{protocol} allows us to find analytical solutions for primal and dual problems, in both deterministic and probabilistic variant, which is not universal property in such kind of problems. 
 
 Please notice, that in deterministic and probabilistic MPBT we use the same symbol for $O_A$ and $X_A$ although they explicit forms are different. However, it is always clear from the context what kind of the protocol we consider.
The solutions of  described below optimisation problems are contained in Section~\ref{mainPBT}, while methods of solving, exploiting representation theory of symmetric group and algebra of partially transposed permutation operators, in Appendices~\ref{PD_prob} and~\ref{AppB}.
\vspace{0.3cm}
\begin{tcolorbox}
{\bf \textit{Primal problem for probabilistic scheme:}}  
The goal is to maximising the average success probability:
	\be
	\label{primalp}
	p^*=\frac{1}{d^{N+k}}\sum_{\mathbf{i}\in\mathcal{I}}\tr \Pi_{\mathbf{i}}=\frac{1}{d^{N+k}}\sum_{\mathbf{i}\in\mathcal{I}}\tr \Theta_{\overline{A}_{\mathbf{i}}},
	\ee
	with $\Pi_{\mathbf{i}}=P^+_{A_{\mathbf{i}}\widetilde{B}}\ot \Theta_{\overline{A}_{\mathbf{i}}}$, subject to
	\be
	\label{sub}
	(a)\quad \forall \mathbf{i}\in\mathcal{I} \quad \Theta_{\overline{A}_{\mathbf{i}}}\geq 0, \qquad (b)\quad \sum_{\mathbf{i}\in\mathcal{I}}P^+_{A_{\mathbf{i}}\widetilde{B}}\ot \Theta_{\overline{A}_{\mathbf{i}}}\leq X_A \ot \mathbf{1}_{B}.
	\ee
	In the above $X_A=O_A^{\dagger}O_A\geq 0$, where $O_A$ is an operation applied by Alice, on  her half of the resource state, satisfying $\tr O_A^{\dagger}O_A=d^N$ (see Figure~\ref{fig1}). The operator $\Theta_{\overline{A}_{\mathbf{i}}}$ acts on $N-k$ systems except those on positions $\mathbf{i}=(i_1,i_2,\ldots,i_k)$.\\
\vspace{0.3cm}
{\bf \textit{Dual problem for probabilistic scheme:}} In order to solve the dual problem, we have to minimise:
\be
\label{pdual}
p_*=d^Nb,\quad \text{where}\quad b\in\mathbb{R}_+,
\ee
subject to
\be
\label{sub2}
(a)\quad \Omega \geq  0, \qquad (b)\quad \tr_{A_{\mathbf{i}}\widetilde{B}}\left(P^+_{A_{\mathbf{i}}\widetilde{B}}\Omega \right)\geq \mathbf{1}_{N-k}, \qquad (c)\quad b\mathbf{1}_N-\frac{1}{d^{N+k}}\tr_{A_{\mathbf{i}}}\Omega \geq 0,
\ee
where $\mathbf{1}_{N-k},\mathbf{1}_N$ denotes the identity operators acting on $N-k$ and $N$ systems respectively, and operator $\Omega$ acts on $N+1$ systems.
\end{tcolorbox}
\vspace{0.5cm}
\begin{tcolorbox}
{\bf \textit{Primal problem for deterministic scheme:}}  The goal is to evaluate the following quantity
\be
\label{primF}
F^*=\frac{1}{d^{2k}}\max_{\{\Pi_{\mathbf{i}}\}}\sum_{\mathbf{i}\in\mathcal{I}}\tr\left(\Pi_{\mathbf{i}}\sigma_{\mathbf{i}} \right),
\ee
with respect to constraints
\be
\label{ccoo}
(a)\quad \forall \mathbf{i}\in\mathcal{I} \quad \Pi_{\mathbf{i}}\geq 0\qquad (b)\quad \sum_{\mathbf{i}\in \mathcal{I}}\Pi_{\mathbf{i}}\leq X_{A}\ot \mathbf{1}_{B},\qquad (c)\quad \tr X_{A}=d^N.
\ee
The matrix $X_{A}$ equals to $O_A^{\dagger}O_A$, where $O_A$ is an operation applied by Alice to her part of the shared entangled pairs (see Figure~\ref{fig1}).

{\bf{\textit{Dual problem for deterministic scheme:}}} The dual problem is to evaluate quantity
\be
\label{Fdual}
F_{*}=d^{N-2k}\min\limits_{\Omega}||\tr_{(k)}\Omega||_{\infty}
\ee
with respect to constraints
\be
\label{cons_dual}
\Omega-\sigma_{\mathbf{i}}\geq 0,\qquad \mathbf{i}\in \mathcal{I}.
\ee
The operator $\Omega$ acts on $N$ systems, $\tr_{(k)}$ denotes partial trace over last $k$ particles, and $\sigma_{\mathbf{i}}$ are the signals states given through~\eqref{sigma}. The symbol $||\cdot||_{\infty}$ denotes the infinity norm.
\end{tcolorbox}

\section{Connection with the algebra of partially transposed permutation operators and efficiency of non-optimal schemes}
\label{connA}
The connection between the ordinary PBT, its multi-port extension and algebra of partially transposed permutation operators has been noticed earlier and exploited for their characterisation in the following cycle of papers~\cite{Studzinski2017,StuNJP,stud2020A}.  Here, for the reader's convenience we  present all the facts necessary for understanding  the results presented further in this manuscript. For the full analysis we refer reader to paper cited in this section.

First of all, let us remind here that the operator $\rho$ from~\eqref{PBT1}, as well as being invariant with respect to action of elements from $S(n-k)$ is also invariant with respect to action of $U^{\ot (n-k)}\ot \overline{U}^{\ot k}$. The bar denotes the complex wise conjugation and $U$ is an element of unitary group $\mathcal{U}(d)$. The same hold for signals $\sigma_{\mathbf{i}}$, where $\mathbf{i} \in \mathcal{I}$, given in~\eqref{sigma}.
This specific kind of symmetry implies that all the basic elements describing considered here teleportation protocols, like signals, MPBT operator, and  measurements~\cite{majenz,2020arXiv200811194L} are elements of the algebra  of partially transposed permutation operators $\mathcal{A}^{(k)}_n(d)$ with respect to $k$ last subsystems~\cite{Stu1,Moz1,MozJPA}. 

\begin{figure}[h]
\centering
	\includegraphics[width=.8\linewidth]{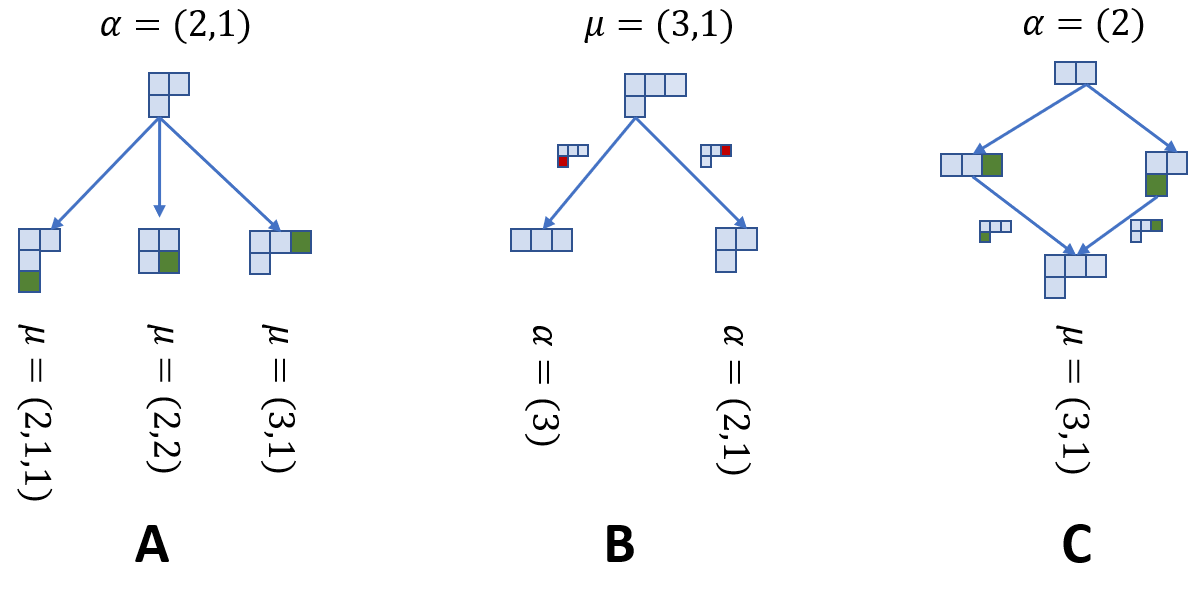}
	\caption{Panel A presents all possible Young frames $\mu \vdash 4$ satisfying relation $\mu \in \alpha$ for $\alpha=(2,1)$ and $k=1$. By the green squares we depict boxes added to initial frame $\alpha$. Panel B presents all possible Young frames $\alpha \vdash 3$ satisfying relation $\alpha \in \mu$ for $\mu=(3,1)$. By red colour we depict boxes subtracted from the initial Young frame $\mu=(3,1)$.    Panel C presents different ways of adding 2 boxes to initial frame $\alpha=(2)$ still getting at the end the same frame $\mu=(3,1)$. In this particular case $m_{(3,1)/(2)}=2$, since the first path is through the frame $(3)$ and the second one through the frame $(2,1)$.}
	\label{relacje_YF}
\end{figure}

To describe the above connection more quantitatively we need a few additional concepts. Namely, a partition $\alpha$ of a natural number $n$ (denoted as $\alpha \vdash n$) is a sequence of positive numbers $\alpha=(\alpha_1,\alpha_2,\ldots,\alpha_r)$, such that $\alpha_1\geq \alpha_2\geq \cdots \geq \alpha_r$ and $\sum_{i=1}^r\alpha_i=n$. Partitions can be represented as a collection of boxes arranged in left-justified rows, we call this a Young frame, and we assign the number $h(\alpha)$, called a height of the frame, equal to the number of its rows. It turns out there is one to one correspondence between partitions (Young frames) and irreducible representations (irreps) for symmetric group $S(n)$~\cite{Fulton1991-book-rep}. More precisely, the number of Young frames for fixed $n$ tells us how many inequivalent irreps $S(n)$ contains. Writing $\mu \vdash n$ and $\alpha \vdash n-k$, we denote the corresponding Young frames, or equivalently irreps of $S(n)$ and $S(n-k)$ respectively. If a frame $\mu \vdash n$ can be obtained from a frame $\alpha \vdash n-k$ by adding $k$ boxes in such a way in every step obtained shape is a valid Young frame, we write $\mu \in \alpha$. Similarly, we denote by the symbol $\alpha \in \mu$ Young frames $\alpha$ which can be obtained from Young frame $\mu$ by subtraction of $k$ boxes in such a way in every step obtained shape is a valid Young frame (see panels A,B in Figure~\ref{relacje_YF}). For any Young frames $\mu \vdash n$ and $\alpha \vdash n-k$ for which $\mu \in \alpha$ there are multiple ways of adding $k$ boxes in a proper way. The number of all possibilities we denote as $m_{\mu/\alpha}$ (see panel C in Figure~\ref{relacje_YF}).
The explicit formula for $m_{\mu/\alpha}$ is not known in general situation and we have to use numerical methods. However, for qubit case ($d=2$) numbers $m_{\mu/\alpha}$ can be expressed due  to results in \cite{aitken_1943,xxx,adin2014enumeration} as the following determinant
\be
m_{\mu/\alpha} = k!\operatorname{det}\begin{pmatrix}
\frac{1}{\alpha_1-\mu_1} && \frac{1}{\alpha_1-\mu_2+1}\\
\frac{1}{\alpha_2-\mu_1-1} && \frac{1}{\alpha_2-\mu_2}
\end{pmatrix},
\ee
where $\alpha_i,\mu_j$, for $1\leq i,j\leq 2$ denote respective rows of Young frames $\alpha$ and $\mu$.

Having the introductory part we can present explicit connection between the most important objects for our further analysis and the algebra of partially transposed permutation operators $\mathcal{A}^{(k)}_d(n)$. It turns out that projectors on irreducible components of $\mathcal{A}^{(k)}_d(n)$, denoted as $F_{\mu}(\alpha)$, where $\alpha \in \mu$,  are also eigenprojectors for MPBT operator $\rho$ from expression~\eqref{PBT1} (see Theorem 17 in~\cite{stud2020A}): 
\be
\label{eigenrho}
\rho=\sum_{\alpha}\sum_{\mu \in \alpha}\lambda_{\mu}(\alpha)F_{\mu}(\alpha).
\ee
Projectors $F_{\mu}(\alpha)$ satisfy the following orthogonality  relations (see Lemma 16 in~\cite{stud2020A}):
\be
F_{\mu}(\alpha)F_{\nu}(\beta)=\delta_{\alpha \beta}\delta_{\mu \nu}F_{\nu}(\beta), \qquad \tr F_{\mu}(\alpha)=m_{\mu/\alpha}m_{\alpha}d_{\mu},
\ee
where $m_{\alpha}$ denotes multiplicity of irrep of $S(N-k)$ labelled by $\alpha$, while $d_{\mu}$ denotes dimension of irrep of $S(N)$ labelled by $\mu$. The numbers $\lambda_{\mu}(\alpha)$ are eigenvalues of $\rho$ and they are equal to (see Theorem 14 and Theorem 17 in~\cite{stud2020A}:
\be
\label{eig}
\lambda_{\mu}(\alpha)=\frac{k!\binom{N}{k}}{d^N}\frac{m_{\mu}}{m_{\alpha}}\frac{d_{\alpha}}{d_{\mu}}.
\ee
Here again $m_{\alpha}, m_{\mu}$ denote respective multiplicities of irreps labelled by $\alpha, \mu$, while $d_{\alpha},d_{\mu}$ stand for their dimensions.

The mathematical tools briefly discussed above and introduced in~\cite{stud2020A} allowed authors in the same paper to  calculate the efficiency of the deterministic and probabilistic protocol in its non-optimal version (see Theorem 22 and Theorem 23 in~\cite{stud2020A}), i.e. when $O_A=\mathbf{1}_A$ in~\eqref{resource0}. Since later on we compare to those results for the reader's convenience we re-state these theorems here with slightly different form more suited for this manuscript.
\begin{theorem}[Theorem 22 in~\cite{stud2020A}]
	\label{Fthm}
	The entanglement fidelity in the deterministic multiport teleportation with $N$ ports and local dimension $d$ is given as
	\be
	\label{Feq1}
	F=\frac{1}{d^{N+2k}}\sum_{\alpha \vdash N-k}\left(\sum_{\mu\in\alpha}m_{\mu/\alpha} \sqrt{m_{\mu}d_{\mu}}\right)^2,
	\ee
	where $m_{\mu},d_{\mu}$ denote multiplicity and dimension of irreducible representations of $S(N)$ respectively, and $m_{\mu/\alpha}$ denotes number of ways in which diagram $\mu$ can be obtained from diagram $\alpha$ by adding $k$ boxes. The measurements are the square-root measurements (SRM) of the form:
	\begin{equation}
	 \forall \ \mathbf{i}\in\mathcal{I}\qquad \Pi_{\mathbf{i}}^{AB}=\frac{1}{\sqrt{\rho}}\sigma^{AB}_{\mathbf{i}}\frac{1}{\sqrt{\rho}},
	\end{equation}
	where the operator $\rho$ is given in~\eqref{PBT1} and the signal states $\sigma^{AB}_{\mathbf{i}}$ in~\eqref{sigma}.
\end{theorem}

\begin{theorem}[Theorem 23 in~\cite{stud2020A}]
\label{thm_p}
The average probability of success in the probabilistic multiport teleportation with $N$ ports and local dimension $d$ is given as
\be
\label{exact}
p=\frac{k!\binom{N}{k}}{d^{2N}}\sum_{\alpha \vdash N-k}\mathop{\operatorname{min}}\limits_{\mu\in\alpha}\frac{m_{\alpha}d_{\alpha}}{\lambda_{\mu}(\alpha)},
\ee
with optimal measurements of the form
\be
\label{ex_measurements}
\forall \ \mathbf{i}\in\mathcal{I}\qquad \Pi_{\mathbf{i}}^{AB}=\frac{k!\binom{N}{k}}{d^{2N}}P^+_{A_{\mathbf{i}}\widetilde{B}}\ot \sum_{\alpha \vdash N-k}P_{\alpha}\mathop{\operatorname{min}}\limits_{\mu\in\alpha}\frac{1}{\lambda_{\mu}(\alpha)}.
\ee
The numbers $\lambda_{\mu}(\alpha)$ are eigenvalues of $\rho$ and are given in~\eqref{eig} and $m_{\alpha}, d_{\alpha}$ denote multiplicity and dimension of the irrep labelled by $\alpha$.
\end{theorem}

	\section{Main results}
	\label{mainPBT}
	In order to find the optimal value for the probability of success in the optimal probabilistic  MPBT scheme, as well entanglement fidelity in its deterministic version,  we have to solve both, the primal and the dual problem and use group theoretic results presented in Section~\ref{factsG}. As we show later, the solutions for the primal and the dual problems coincide, giving us the real value of probability of success and entanglement fidelity. From the general structure of the proofs we are also able to extract explicit form of the optimal sets of measurements and rotations $O_A$ applied by Alice. 
	\subsection{Optimal probabilistic MPBT scheme}
	We start the main result of this section, describing probability of success in teleportation process:
	\begin{theorem}
	\label{Thmp}
	The averaged probability of success $p$ for the optimal probabilistic MPBT scheme, with $N$ ports, each of dimension $d$, while teleporting $k$ systems, is given by the following expression:
	\be
	\label{p}
	p_{succ}=\frac{N!}{(N-k)!}\frac{(d^2+N-k-1)!}{(d^2+N-1)!}.
	\ee
\end{theorem}
\begin{figure}[h]
\centering
	\includegraphics[width=\linewidth]{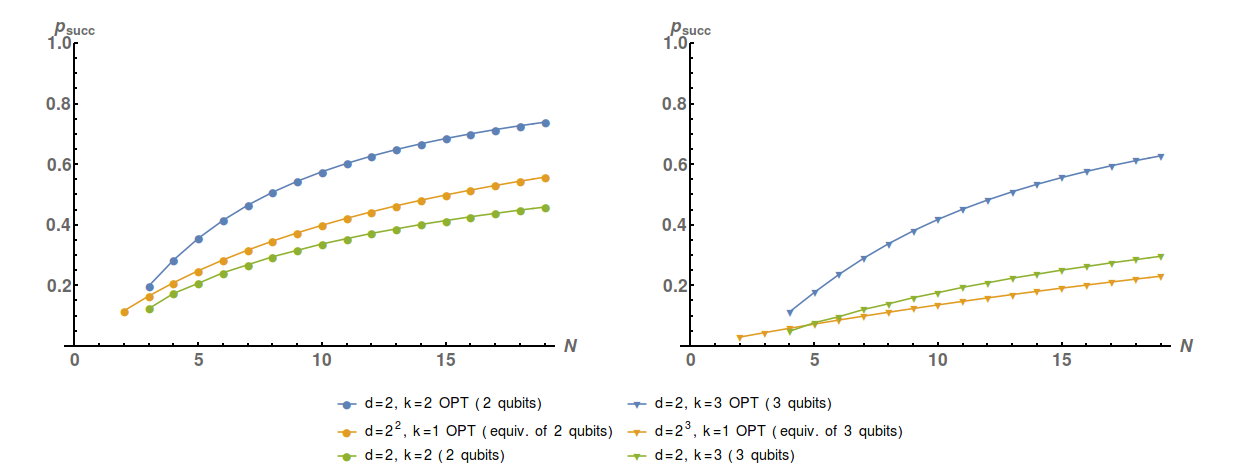}
	\caption{Figure presents  comparison of the optimal multi-port probabilistic scheme ($k>1$, OPT)  with its non-optimal version. In all variants the efficiency is substantially higher. Additionally, we plot probability for the standard optimal PBT, for $d=4$ and $d=8$. We see that our scheme for $k=2, d=2$ performs better, which was not the case in the non-optimal multi-port teleportation procedure for $k=2, d=2$.}
	\label{fig2a}
\end{figure}
The proof of this theorem, in a form of solution of a SDP problem, is located in Appendix~\ref{PD_prob}.
The explicit values of averaged success probability as a function of ports, while teleporting different number of particles, are plotted in Figure~\ref{fig2a}. We see in every case the optimal protocol performs better than the non-optimal MPBT.
We see, that for $k=1$, expression~\eqref{p} reduces to the average probability of success in the standard probabilistic PBT scheme~\cite{Studzinski2017}, and gives
\be
p_{succ}=\frac{N}{d^2+N-1}=1-\frac{d^2-1}{d^2+N-1}.
\ee 
The limit of $p_{succ}$ from~\eqref{p}, when teleporting composed state of  $k$ subsystems, when $k$ is independent on $N$, is $\lim_{N\rightarrow\infty}=1$, since we can write this expression as
\be
	p_{succ}={N \choose k} {N + d^2 - 1 \choose k}^{-1}.
\ee
Now, one could ask if high average probability of teleportation possible when the number of teleported systems increases with number of ports. Namely, we want to examine the limit when $k$ depends on $N$, i.e. $k= k(N)$. We summarise our findings in the following theorem.
\begin{theorem}
\label{kN}
For the probabilistic MPBT scheme with $N$ ports of dimension $d$ each, whenever the number $k=k(N)$ of teleported systems changes as $o(N)$ the probability of success reaches 1 with $N\rightarrow \infty$.
\end{theorem}
The proof of this theorem is located in Appendix~\ref{DkN}. Restricting ourselves to qubits $(d=2)$, we get a square improvements with respect to non-optimal version described and analysed in~\cite{Kopszak2020}. Namely, from $k=o(\sqrt{N})$ we improve to $k=o(N)$. The asymptotic behaviour of the optimal probabilistic MPBT compared to the optimal probabilistic PBT protocol offers a qualitative improvement. Namely, the latter allows for faithful teleportation (i.e. $p_{succ}$ goes to 1) when $N\to \infty$ only when $k=o(N^{1/2})$, however for the former one this is achieved whenever $k=o(N)$.

Finally, we are in position to present explicit form of the optimal POVMs and the operation $O_A$ performed by Alice (see Figure~\ref{fig1}).
\begin{theorem}
\label{optmeas}
In the optimal probabilistic MPBT, with $N$ ports of dimension $d$ each, the optimal POVMs $\{\Pi_{\mathbf{i}}\}_{\mathbf{i}\in\mathcal{I}}$ and optimal operation $O_A$ are given respectively by the following expressions:
\be
\Pi_{\mathbf{i}}=P^+_{A_{\mathbf{i}}B}\otimes\sum_{\alpha\vdash N-k}\frac{d^{N+k}\frac{m_\alpha}{\sum_{\nu\vdash N} m_\nu^2}}{k!{N \choose k}d_\alpha}P_\alpha,\quad O_A= \sqrt{d^N}\sum_\mu\sqrt{\frac{m_\mu}{d_\mu\sum_{\nu\vdash N} m^2_\nu}} P_\mu.
\ee
By $P_{\alpha},P_{\mu}$ we denote Young projectors on irreps of $S(N-k)$ and $S(N)$ labelled by $\alpha \vdash N-k$ and $\mu \vdash N$ respectively.
\end{theorem}
The proof of this theorem is located at the end of Appendix~\ref{PD_prob}.
\subsection{Optimal deterministic MPBT scheme}
\begin{figure}[h]
\centering
\includegraphics[width=.75\linewidth]{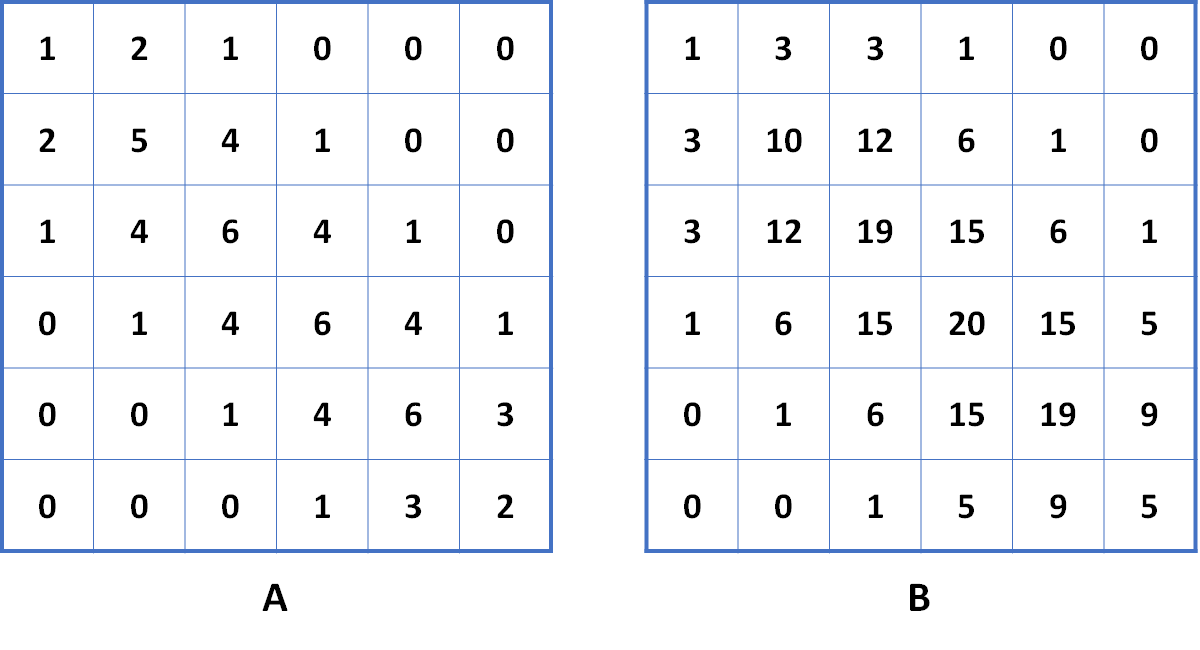}
	\includegraphics[width=.75\linewidth]{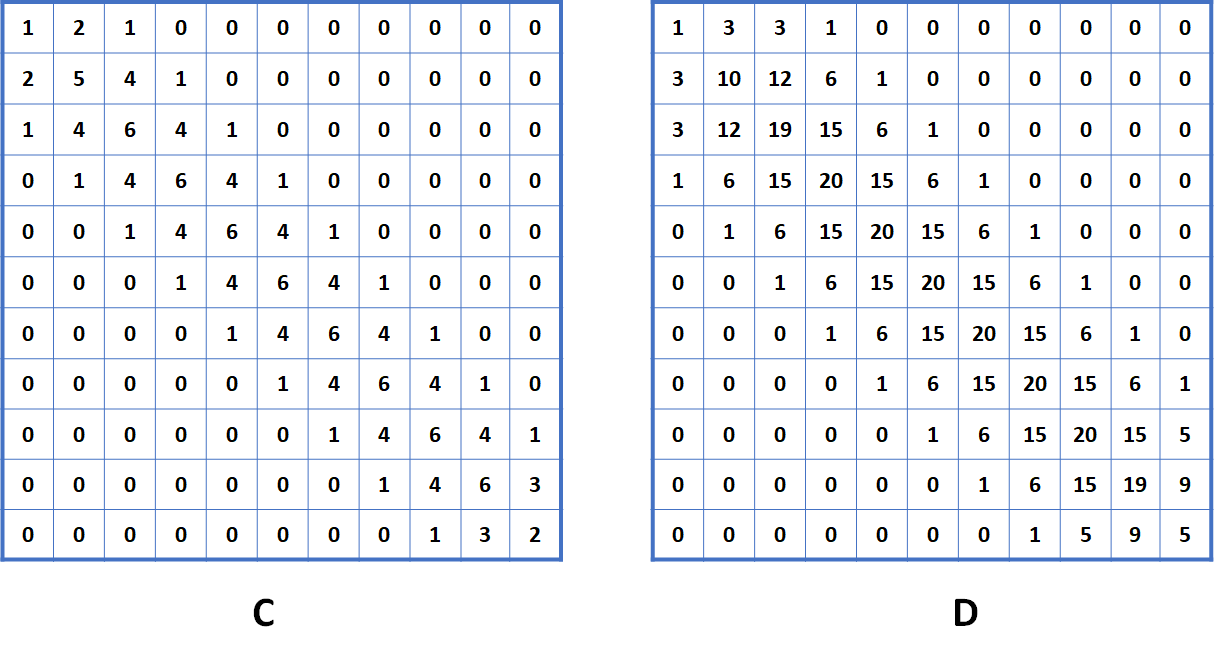}
	\caption{Figure presents  generalised teleportation matrix $M_F^{d,k}$ from Definition~\ref{gen_MF} in qubit case $d=2$. Panel A present matrix for $N=10$ and teleporting $k=2$ systems, while panel B presents matrix also for $N=10$ but teleporting $k=3$ systems. Panel C presents matrix for $N=20$ and teleporting $k=2$ systems, while panel D is for teleporting $k=3$ systems. Notice that increasing number of teleported systems $k$ for fixed number of ports $N$ the teleportation matrix is less sparse. Here, even in the qubit case we are not in the set of matrices for which we know the spectrum analytically - as it was in the qubit case for $k=1$, see Section 5.3 in~\cite{StuNJP}. Despite this, for every $k>1$, still we can effectively evaluate maximal eigenvalue using algorithm described in Section 5.4 in~\cite{StuNJP}.  }
	\label{MFN10}
\end{figure}

\begin{figure}[h]
	\includegraphics[width=\linewidth]{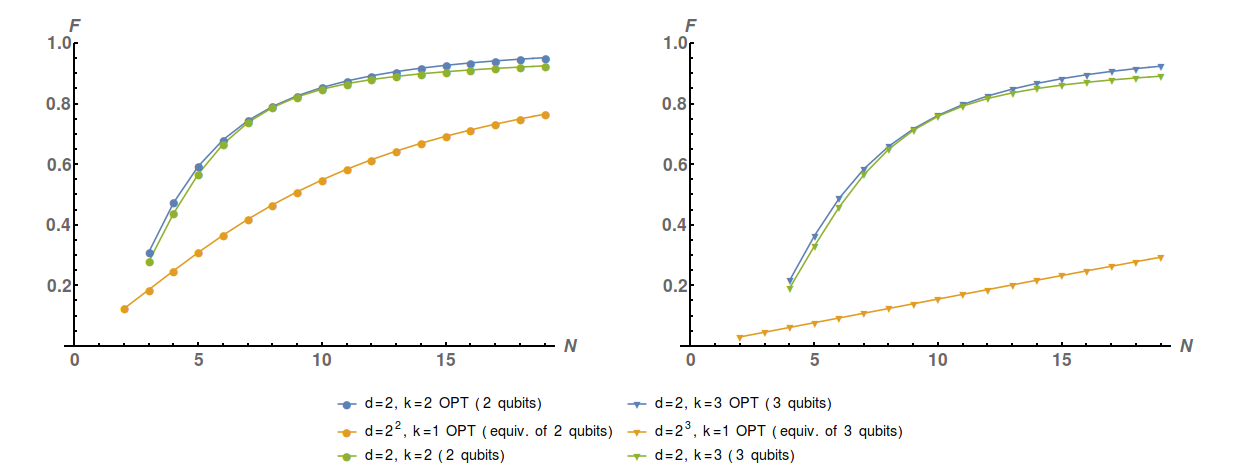}
	\caption{Figure presents comparison of the optimal deterministic MPBT scheme ($k>1$, OPT) with its non-optimal version. In every case the efficiency in the optimal case is higher. Additionally, to illustrate the efficiency jump, we plot entanglement fidelities for the standard optimal PBT ($k=1$).}
	\label{fig2}
\end{figure}

In the second part of this section we present results concerning the entanglement fidelity in the optimal deterministic MPBT scheme, similarly as it was for the probabilistic scheme, by solving respective primal and dual problem. We show that solutions coming from both of them coincide, giving us the optimal value. Additionally, we present a form of the optimal set of measurements and optimal rotation $O_A$. Before we proceed, we have to introduce concept of  generalised teleportation matrix:
\begin{definition}
	\label{gen_MF}
	Let $\mu,\nu$ run over all possible irreps of the permutation group $S(N)$ and $\alpha$ numbers irreps of $S(N-1)$, whose height is less or equal $d$, then we define matrix $M_F^{d,k}$ as
	\be
	(M_F^{d,k})_{\mu\nu}\equiv \sum_{\alpha \in \mu}m^2_{\mu/\alpha}\delta_{\mu\nu}+(1-\delta_{\mu\nu})\Delta_{\mu\nu},
	\ee
	with
	\be
	\Delta_{\mu\nu}\equiv \begin{cases} \sum_{\alpha \in \mu \wedge \nu}m_{\mu/\alpha}m_{\nu/\alpha} \quad \text{if} \quad \mu \sim_k \nu,\\
		0 \qquad \qquad  \qquad \qquad \text{otherwise}.
	\end{cases}
	\ee
The symbol $\mu \sim_k \nu$ means that the frame $\nu$ can be obtained from the frame $\mu$ by moving up to $k$ boxes one by one. The symbol $m_{\mu/\alpha}$ denotes number of ways in which a diagram $\mu$ can be obtained from a diagram $\alpha$ by adding $k$ boxes one by one. Lastly, the symbol $\alpha \in \mu \wedge \nu$ means Young frame $\alpha$ can be obtained from Young frame $\mu$ as well as from $\nu$ by subtracting $k$ boxes.
\end{definition}
Arranging matrix entries of the generalised teleportation matrix we assume that all indices $\mu,\nu$ are ordered in the strongly
decreasing lexicographic order, starting from the biggest Young diagram $\mu=(N)$. In such ordering, Young
diagrams strongly decrease, whereas the height of the Young diagrams weakly increases. In Figure~\ref{MFN10} we show explicit form of matrix $M_F^{d,k}$, for clarity we restrict ourselves to qubit case $(d=2)$.
Definition~\ref{gen_MF} generalises the concept of the teleportation matrix introduced in~\cite{StuNJP}. Indeed, having $k=1$ we have $\alpha \vdash N-1$ and $\mu,\nu \vdash N$. In this case we always have $m_{\mu/\alpha}\in \{0,1\}$, since diagrams $\mu$ can obtained form $\alpha$ by adding a single box if $\mu \in \alpha$. It means that the total number of paths from $\alpha$ to $\mu$ can be equal at most 1.
From the similar reasons the symbol $\Delta_{\mu \nu}$ can take only two values: 1 if $\mu \sim_1 \nu$ or 0 otherwise. Having  Definition~\ref{gen_MF} we show how the generalised teleportation matrix is connected with entanglement fidelity. Namely, we have the following:
\begin{theorem}
\label{DetF}
The entanglement fidelity for optimal deterministic MPBT teleportation scheme, with $N$ ports of dimension $d$ each, is given by
\be
F=\frac{1}{d^{2k}}\lambda_{\max}(M_F^{d,k}),
\ee
where $k$ is the number of teleported systems and $\lambda_{\max}(M_F^{d,k})$ is maximal eigenvalue of generalised teleportation matrix $M_F^{d,k}$ from Definition~\ref{gen_MF}.
\end{theorem}
The proof of this theorem is located in Appendix~\ref{AppB}. Taking $k=1$ we reproduce expression for entanglement fidelity in the ordinary optimal port-based teleportation, see~\cite{StuNJP}. In Figure~\ref{fig2} we present plots of functional dependence of entanglement fidelity  as a function of ports, while teleporting different number of systems.
Similarly, as it was for the optimal probabilistic version, here we also present the explicit form of optimal measurements and operation $O_A$ applied by Alice to her part of shared maximally entangled pairs. The proof of the following theorem is located in Appendix~\ref{AppB}.
\begin{theorem}
\label{optmeasurements}
In the optimal deterministic MPBT, with $N$ ports of dimension $d$ each, the optimal POVMs $\{\Pi_{\mathbf{i}}\}_{\mathbf{i}\in\mathcal{I}}$ are given by
\be
\Pi_{\mathbf{i}}=\Pi \sigma_{\mathbf{i}} \Pi\quad \text{with}\quad \Pi=\sum_{\alpha \vdash N-k}\sum_{\mu\in \alpha}\frac{d_{\mu}}{\sqrt{k!{N\choose k}}}\sqrt{\frac{m_\alpha}{d_\alpha}}\frac{v_\mu}{m_\mu}F_\mu(\alpha),
\ee
where $F_\mu(\alpha)$ are the eigen-projectors from expression~\eqref{eigenrho}.
Optimal operation $O_A$ has a form
 \begin{equation}
    O_A= \sqrt{d^N}\sum_\mu\frac{v_\mu}{d_\mu m_\mu} P_\mu,
    \end{equation}
where $P_{\mu}$ denotes Young projector on irreps of $S(N)$ labelled by $\mu\vdash N$.
In both equations the numbers $v_\mu>0$ are the components of an eigenvector corresponding to the greatest eigenvalue of the generalised teleportation matrix $M_F^{d,k}$. 
\end{theorem}

\section{New results concerning representation theory}
\label{factsG}
In this section we formulate new results in group theory of the symmetric group which are necessary to get closed expression for the average probability of success in Theorem~\ref{Thmp}, as well as, for the formulation of Theorem~\ref{kN}. Theorem connects squared multiplicities  of the symmetric groups $S(N)$ and $S(N-k)$ with global parameters, such as local dimension $d$ and number $N$.
Lastly, we formulate  lemma concerning calculating trace in algebra of partially transposed permutation operators allowing to obtain connection between entanglement fidelity and the generalised teleportation matrix, see Theorem~\ref{DetF}.

Let us start from reminding the standard swap representation of symmetric group $S(N)$
\be
\label{swap}
V_{N}^{d}:S(N)\rightarrow \operatorname{Hom}[(\mathbb{C}^{d})^{\otimes N}]. 
\ee
In other words, for any $\pi \in S(N)$,  element $V_N^d(\pi)=V(\pi)$ permutes factors in the space $\mathcal{H\equiv (\mathbb{C}}^{d})^{\otimes n}$:
\be
\forall \pi \in S(n)\qquad V(\pi )|e_{1}\>\otimes |e_{2}\>\otimes
\cdots \otimes |e_{n}\>=|e_{\pi ^{-1}(1)}\>\otimes |e_{\pi
^{-1}(2)}\>\otimes \cdots \otimes |e_{\pi ^{-1}(n)}\>,
\ee
where  $\{|e_{i}\>\}_{i=1}^{d}$ is an orthonormal basis of the space $\mathcal{\mathbb{C}}^{d}.$
It is well-known fact~\cite{Fulton1991-book-rep} that it reduces to the form 
\be
\label{decomp}
V_{N}^{d}\cong \bigoplus _{\nu :h(\nu )\leq d}m_{\nu }\psi ^{\nu }, 
\ee
where $\psi^{\nu }$ are the irreps of $S(N)$, $m_{\nu }$ are multiplicities of $\psi^{\nu }$. The direct sum runs over all Young diagrams $\nu \vdash N$ with height $h(\nu)\leq d$. Having this definition we formulate the first main result of this section:
\begin{theorem}
	\label{thm}
	Let us consider decomposition into irreps the swap representations $V_{N}^{d}$ and $V_{N-k}^{d}$:
	\be
	V_{N}^{d}\cong \bigoplus _{\nu :h(\nu )\leq d}m_{\nu }\psi ^{\nu },\qquad V_{N-k}^{d}\cong \bigoplus _{\alpha :h(\alpha )\leq
		d}m_{\alpha }\varphi ^{\alpha },
	\ee
	then  we have 
	\be
	\frac{\sum_{\nu :h(\nu )\leq d}m_{\nu }^{2}}{\sum_{\alpha :h(\alpha )\leq
			d}m_{\alpha }^{2}}=\frac{d^{2}+N-1}{N}\frac{d^{2}+N-2}{N-1}\cdots \frac{%
		d^{2}+N-k}{N-k+1}.
	\ee
\end{theorem}
The proof of the above theorem is contained in Appendix~\ref{PT7}.
Finally, we formulate the second main result for this section, namely lemma concerning calculation of trace in the algebra of the partially transposed permutation operators. It involves eigen-projectors $F_{\mu}(\alpha)$ from~\eqref{eigenrho} and canonical transposed permutation operator $V^{(k)}$ from~\eqref{canonical}.
\begin{lemma}
\label{coef}
For fixed $1\leq l\leq m_{\alpha}$ by $\{|\varphi_{k,l}(\alpha)\>\}_{k=1}^{d_{\alpha}}$ we denote set of basis vectors in irrep $\alpha \vdash N-k$. Then for every projector $F_{\mu}(\alpha)$ and partially transposed permutation operator $V^{(k)}$ the following relation holds:
\be
\forall \ \mu\in\alpha \quad \tr\left[|\varphi_{k,l}(\alpha)\>\<\varphi_{r,s}(\alpha)|V^{(k)}F_{\mu}(\alpha)\right]=m_{\mu/\alpha}\frac{m_{\mu}}{m_{\alpha}}\delta_{kr}\delta_{ls}.
\ee
\end{lemma}
The proof of the above lemma is contained in Appendix~\ref{PL8}.

\section{Conclusions}
In this paper we formulate the optimal version of multi-port based teleportation scheme. This kind of protocol allows for teleporting a number of unknown states or one composite system in one go, with higher efficiency than original variants of PBT and non-optimal MPBT. We discuss probabilistic and deterministic variant of the protocol. In the first case we derive expression for the average probability of success in remarkably compact form. Namely, efficiency depends only on number of ports $N$, their dimension $d$, and number $k$ of teleported systems. We get  significant improvement in the efficiency with respect to non-optimal MPBT . Since in the general situation, the number $k$ of teleported systems can vary with number of ports, we show that to get $p=1$  asymptotically, one can take at most $k(N)=o(N)$ systems to be teleported. In qubit case it proves that we  get square improvement with respect to non-optimal MPBT, although numeric shows we perform better for any $d>2$. Finally, we derive the explicit form of operation $O_A$ applied by Alice to her part of ports and form of optimal POVMs. In case of deterministic scheme we prove that entanglement fidelity of the whole process can be expressed in term of maximal eigenvalue of generalised teleportation matrix, also introduced in this manuscript. This connection allows us to show that indeed we perform better than in non-optimal MPBT. The set of optimal POVMs and operation $O_A$ is also presented. All derived results require novel results from representation theory of the symmetric group and recently studied algebra of partially transposed permutation operators $\mathcal{A}^{(k)}_n(d)$. In particular we derive a connection between sum of squared multiplicities of irreducible representations of symmetric groups $S(N)$ and $S(N-k)$, which is up  to our best knowledge not known and can be of separate interest. From the side of the algebra $\mathcal{A}^{(k)}_n(d)$, we prove lemma allowing us performing effective computations involving its irreducible projectors. 

We leave here a few open questions for further possible research regarding optimal protocols. As we see, in the optimal deterministic MPBT fidelity of the whole process is given as a function of maximal eigenvalue of the generalised teleportation matrix $M_F^{d,k}$. Except one specific case, for $d=2$ and $k=1$ discussed in~\cite{StuNJP}, eigenvalues of  $M_F^{d,k}$ can evaluated only numerically. The natural question is there any method for bounding possibly sharply the maximal eigenvalue to have analytical bound for the fidelity, or at least bound in terms of group-theoretic quantities, like irreps dimensions and multiplicities. Solving this particular problem we could later find what is the precise scaling of fidelity in $N$ and $d$, similarly as it was done in~\cite{majenz}. The problem of scaling for an arbitrary $d$ and $N$ is open also for non-optimal MPBT (except qubits~\cite{Kopszak2020}).

\section*{Acknowledgements}
MS, PK and MM are supported through grant Sonatina 2, UMO-2018/28/C/ST2/00004 from the Polish National Science Centre. 
\appendix 
\section{Solution of primal and dual problem for probabilistic scheme}
\label{PD_prob}
{\bf Solving the primal problem.}	 
In the probabilistic scheme we have faithful teleportation, meaning that the entanglement fidelity between the input and the output state is $F=1$. Now, denoting by $P^+_{CD}\cong P^+_{C_1D_1}\ot \cdots \ot P^+_{C_kD_k}$ maximally entangled state, where a half, i.e. systems $C_1,\ldots, C_k$, are teleported to Bob, we demand the following constraint to be satisfied by Alice's POVMs $\{\Pi_{\mathbf{i}}\}_{_{\mathbf{i}}\in\mathcal{I}}$:
\be
\forall \ \mathbf{i}\in \mathcal{I} \qquad \tr\left[\Pi_{\mathbf{i}}\left(\mathbf{1}-P^+_{A_{\mathbf{i}}B}\right)\right]=0.
\ee
This constraint implies that measurements must have a form
\be
\label{gen1}
\forall \ \mathbf{i}\in \mathcal{I} \qquad \Pi_{\mathbf{i}}=P^+_{A_{\mathbf{i}}B}\otimes \Theta_{\overline{A}_{\mathbf{i}}},
\ee
where $\Theta_{\overline{A}_{\mathbf{i}}}$ are arbitrary operators acting on all port systems, except those which were projected on maximally entangled pairs $P^+_{A_{\mathbf{i}}B}$.
Our task in this section is to determine an explicit form of $\Theta_{\overline{A}_{\mathbf{i}}}$, giving maximal possible average probability of success. Let us assume the following form of the operators $\Theta_{\overline{A}_{\mathbf{i}}}$ and $X_A$:
	\be
	\label{gen2}
	\Theta_{\overline{A}_{\mathbf{i}}}:=\sum_{\alpha \vdash N-k}u(\alpha)P_{\alpha}^{\overline{\mathbf{i}}},\qquad X_A:=\sum_{\mu \vdash N}c_{\mu}P_{\mu},
	\ee
	where $c_{\mu}\geq 0$ for all $\mu \vdash N$.
	Notation $P_{\alpha}^{\overline{\mathbf{i}}}$ means a Young projectors, in irrep $\alpha$ of $S(N-k)$,  acting on $N-k$ systems, except those which were projected on maximally entangled pairs $P^+_{A_{\mathbf{i}}B}$. It means that for every multi index $\mathbf{i}$, we consider the same projector $P_{\alpha}$, but supported on different subsystems. By $P_{\mu}$ we denote a Young projector acting on irrep $\mu\vdash N$ of $S(N)$. Please notice that the form of $X_A=O_A^{\dagger}O_A$ in~\eqref{gen2} encodes the explicit form of the operation $O_A$ applied by Alice. Indeed, taking into account property $O_A^{\dagger}=O_A$ and the fact that projectors $P_{\mu}$ have orthogonal supports for different $\mu$, we can write $O_A$ as
	\begin{equation}
	O_A=\sum_{\mu \vdash N}\sqrt{c_{\mu}}P_{\mu}.
	\end{equation}
	We assume that the numbers $u(\alpha)$ and $c_{\mu}$ are given by the following expressions:
	\be
	\label{u}
	u(\alpha):=\frac{d^{N+k}g(N)m_{\alpha}}{k!\binom{N}{k}d_{\alpha}},\quad \text{with}\quad g(N):=\frac{1}{\sum_{\nu \vdash N}m^2_{\nu}},
	\ee
	and
	\be
	\label{c}
	\forall \mu\vdash N \qquad c_{\mu}:=\frac{d^Ng(N)m_{\mu}}{d_{\mu}}.
	\ee
	Form~\eqref{u} and~\eqref{c} we deduce the relation
	\be
	\label{uc}
	\forall \ \alpha \vdash N-k \quad \forall \mu\in\alpha \qquad u(\alpha)=\frac{d^k}{\gamma_{\mu}(\alpha)}c_{\mu},
	\ee
	where $\gamma_{\mu}(\alpha)=k!\binom{N}{k}\frac{m_{\mu}d_{\alpha}}{m_{\alpha}d_{\mu}}$ are the unnormalised eigenvalues of the MPBT operator in~\eqref{PBT1}. This explicit form of $\gamma_{\mu}(\alpha)$ has been derived in~\cite{stud2020A}.
	Due to the above choice of the coefficients conditions (a) from~\eqref{sub} are automatically satisfied. Now, we move to the constraint (b) from~\eqref{sub}. Let us expand the left-hand side of (b):
	\be
	\label{uc2}
	\begin{split}
	\sum_{\mathbf{i}\in\mathcal{I}}P^+_{A_{\mathbf{i}}\widetilde{B}}\ot \Theta_{\overline{A}_{\mathbf{i}}}&=\sum_{\tau \in \mathcal{S}_{n,k}} \ \sum_{\alpha \vdash N-k}\frac{u(\alpha)}{d^k} V(\tau^{-1})P_{\alpha} V^{(k)} V(\tau)=\sum_{\alpha \vdash N-k}\frac{u(\alpha)}{d^k}\eta(\alpha)=\sum_{\alpha \vdash N-k}\sum_{\mu \in \alpha}\frac{u(\alpha)}{d^k}\eta_{\mu}(\alpha)\\
	&=\sum_{\alpha \vdash N-k}\sum_{\mu \in \alpha}\frac{u(\alpha)}{d^k}P_{\mu}\eta(\alpha)P_{\mu},
	\end{split}.
\ee
The first equality in~\eqref{uc2} follows from one-to-one correspondence between the set $\mathcal{I}$ of allowed indices $\mathbf{i}$ and the coset $\mathcal{S}_{n,k}$. Namely, we can always find a permutation $\tau$ from $\mathcal{S}_{n,k}$ transforming one POVM to another. This covariance property allows us to take POVM corresponding to canonical index $\mathbf{i}_0$, discussed in Section~\ref{protocol}, and obtain any other effect by applying one of the permutation from $\mathcal{S}_{n,k}$. The third equality follows from the observation that operators $\eta(\alpha)$ which are of the form
\begin{equation}
  \eta(\alpha)=  \sum_{\tau \in \mathcal{S}_{n,k}}V(\tau^{-1})P_{\alpha} V^{(k)} V(\tau)=d^N\rho(\alpha),
\end{equation}
 are proportional to MPBT operators $\rho$ from~\eqref{PBT1} restricted to  irreducible spaces labelled by $\alpha$ of the algebra $\mathcal{A}_n^{(k)}(d)$.  In particular, every operator $\eta(\alpha)$ is invariant with respect to elements from $S(N)$, so it can be written as
 \begin{equation}
 \forall \alpha \vdash N-k \quad \eta(\alpha)=\sum_{\mu\in\alpha} \eta_{\mu}(\alpha)= \sum_{\mu\in\alpha}P_{\mu}\eta(\alpha)P_{\mu},
 \end{equation}
where $P_{\mu}$ is projector on irreducible representation indexed by $\mu \vdash N$ of the permutation group $S(N)$. This allows us to conclude that $\eta_{\mu}(\alpha)$ have the same spectral properties, as the components $\rho_{\mu}(\alpha)$ of the MPBT operator $\rho$, and we can apply results derived in Theorem 17 in~\cite{stud2020A}. Now, to satisfy constraint (b), with coefficients from~\eqref{u},~\eqref{c}, for every $\alpha \vdash N-k$ and $\mu \in \alpha$, we must obey
\be
\frac{u(\alpha)}{d^k}\gamma_{\mu}(\alpha)\leq c_{\mu}.
\ee
Indeed, plugging~\eqref{uc} to left-hand side of~\eqref{uc2} we satisfy the constraint (b) from~\eqref{sub} with equality. Now we are in position to evaluate $p^*$ from~\eqref{primalp}:
\be
\begin{split}
	p^*&=\frac{1}{d^{N+k}}\sum_{\mathbf{i}\in\mathcal{I}}\tr \Theta_{\overline{A}_{\mathbf{i}}}=\frac{1}{d^{N+k}}\sum_{\mathbf{i}\in\mathcal{I}}\sum_{\alpha \vdash N-k} u(\alpha)\tr\left(P_{\alpha}^{\overline{\mathbf{i}}}\right)=\frac{k!\binom{N}{k}}{d^{N+k}} \sum_{\alpha \vdash N-k} u(\alpha)\tr\left(P_{\alpha}\right)\\
	&=\frac{k!\binom{N}{k}}{d^{N+k}} \sum_{\alpha \vdash N-k}\frac{d^{N+k}g(N)m_{\alpha}}{k!\binom{N}{k}d_{\alpha}}m_{\alpha}d_{\alpha}=\frac{\sum_{\alpha \vdash N-k}m_{\alpha}^2}{\sum_{\nu \vdash N}m_{\nu}^2}=\frac{N(N-1)\cdots (N-k+1)}{(d^2+N-1)(d^2+N-2)\cdots (d^2+N-k)}\\
	&=\frac{N!}{(N-k)!}\frac{(d^2+N-k-1)!}{(d^2+N-1)!},
\end{split}
\ee
where in the second line we applied Theorem~\ref{thm}.\\
{\bf Solving the dual problem} From dual problem we can require the same types of symmetries as in the primal one. Due to this, we assume the operator $\Omega$ is chosen as a linear combination of the projectors $F_{\mu}(\alpha)$, so structure analogous to MPBT operator given in~\eqref{eigenrho}:
\be
\label{omega}
\Omega=\sum_{\alpha \vdash N-k}\sum_{\mu \in \alpha}x_{\mu}(\alpha)F_{\mu}(\alpha),\quad \text{where}\quad \forall \alpha \ \forall \mu \in \alpha  \quad  x_{\mu}(\alpha)\geq 0.
\ee
The condition (a) from~\eqref{sub2} is automatically satisfied, because of the assumed form of the operators $\Omega$. Now, let us evaluate the partial traces from expressions (b) and (c) in~\eqref{sub2}. Due to invariance property with respect to elements from the coset $\mathcal{S}_{n,k}$, it is enough to compute the trace for canonical index $\mathbf{i}_0$. The covariance property reduces calculations of different partial traces in (b) and (c) to those defined on last systems. Introducing $\tr_{(2k)}=\tr_{n-2k+1,\ldots,n-k}$ and $\tr_{(k)}=\tr_{n-2k+1,\ldots,n}$, we write:
\be
\label{1}
\begin{split}
\frac{1}{d^k}\tr_{(2k)}\left(V^{(k)}\Omega \right)&=\frac{1}{d^k}\sum_{\alpha \vdash N-k}\sum_{\mu \in \alpha}x_{\mu}(\alpha)\tr_{(2k)}\left(V^{(k)}F_{\mu}(\alpha) \right)=\frac{1}{d^k}\sum_{\alpha \vdash N-k}\sum_{\mu \in \alpha}x_{\mu}(\alpha)\tr_{(2k)}\left(V^{(k)}P_{\alpha}P_{\mu} \right)\\
&=\frac{1}{d^k}\sum_{\alpha \vdash N-k}\sum_{\mu \in \alpha}x_{\mu}(\alpha)\tr_{(k)}\left(P_{\alpha}P_{\mu} \right)=\frac{1}{d^k}\sum_{\alpha \vdash N-k}\sum_{\mu \in \alpha}x_{\mu}(\alpha)m_{\mu/\alpha}\frac{m_{\mu}}{m_{\alpha}}P_{\alpha},
\end{split}
\ee
and
\be
\label{2}
\begin{split}
\tr_{(k)}\Omega=\sum_{\alpha \vdash N-k}\sum_{\mu \in \alpha}x_{\mu}(\alpha)\tr_{(k)}\left(F_{\mu}(\alpha) \right)=\sum_{\alpha \vdash N-k}\sum_{\mu \in \alpha}x_{\mu}(\alpha)m_{\mu/\alpha}\frac{m_{\alpha}}{m_{\mu}}P_{\mu}.
\end{split}
\ee
In both final expressions coefficients $m_{\mu/\alpha}$ tell us in how many ways we can get a Young frame $\mu \vdash N$ from a Young frame $\alpha \vdash N-k$ by adding $k$ boxes in a proper way. In expression~\eqref{1} we use Lemma 18 and Lemma 21 from~\cite{stud2020A}, while in~\eqref{2} we apply Lemma 19 from the same paper.

Now, we have to find optimal values for the coefficients $x_{\mu}(\alpha)$ in the decomposition~\eqref{omega}. We assume they have a following form:
\be
\label{x}
\forall \alpha \ \forall \mu \in \alpha  \quad  x_{\mu}(\alpha)=d^k\frac{p(d,k,N)}{m_{\mu/\alpha}}\frac{m_{\mu}}{m_{\alpha}},
\ee
where the function $p(d,k,N)$ is given through the right-hand side of~\eqref{p}. Let us substitute the explicit form of $x_{\mu}(\alpha)$ to (b) in~\eqref{sub2}, using result in~\eqref{1}:
\be
\begin{split}
\frac{1}{d^k}\tr_{(2k)}\left(V^{(k)}\Omega \right)&=p(d,k,N)\sum_{\alpha \vdash N-k}\sum_{\mu \in \alpha}\frac{m_{\mu}^2}{m_{\alpha}^2}P_{\alpha}.
\end{split}
\ee
Since the identity on the right-hand side of (b) in~\eqref{sub2} can be written as $\mathbf{1}=\sum_{\alpha}P_{\alpha}$, we conclude that the constraint (b) in~\eqref{sub2} is satisfied. Indeed, we have
\be
\forall \  \alpha \vdash N-k \qquad p(d,k,N)\sum_{\mu \in \alpha}\frac{m_{\mu}^2}{m_{\alpha}^2}\geq p(d,k,N)\frac{\sum_{\mu\in\alpha}m_{\mu}^2}{\sum_{\alpha \vdash N-k}m_{\alpha}^2}=1,
\ee
where in the last step we used Theorem~\ref{thm}. Now, substituting~\eqref{x} to~\eqref{2} one can get
\be
\tr_{(k)}\Omega=\sum_{\alpha \vdash N-k}\sum_{\mu \in \alpha}x_{\mu}(\alpha))m_{\mu/\alpha}\frac{m_{\alpha}}{m_{\mu}}P_{\mu}=d^kp(d,k,N)\sum_{\alpha \vdash N-k}\sum_{\mu \in \alpha}P_{\mu}=d^kp(d,k,N)\mathbf{1}_N.
\ee
To satisfy the condition (c) from~\eqref{sub2}, which reads now
\be
\left( b-\frac{1}{d^N}p(d,k,N)\right) \mathbf{1}_N\geq 0,
\ee
we need to take $b=\frac{1}{d^N}p(d,k,N)$. Thus we have then equality in (c) from~\eqref{sub2}, so the value of $b$ is also the minimal possible. This allows us to conclude that $p_*=d^Nb=p(d,k,N)$ in~\eqref{pdual}.
This equality proves the Theorem~\ref{Thmp}.
\vspace{0.3cm}\\
The above considerations allow us to deduce the explicit form of the optimal set of measurements and operation $O_A$ applied by Alice. Indeed, combining~\eqref{gen1} with~\eqref{gen2} and~\eqref{u} we arrive to
\be
\Pi_{\mathbf{i}}=P^+_{A_{\mathbf{i}}B}\otimes\sum_{\alpha\vdash N-k}\frac{d^{N+k}\frac{m_\alpha}{\sum_{\nu\vdash N} m_\nu^2}}{k!{N \choose k}d_\alpha}P_\alpha.
\ee
For the optimal operation $O_A$ it is enough to plug~\eqref{c} to~\eqref{gen2}:
\be
O_A= \sqrt{d^N}\sum_\mu\sqrt{\frac{m_\mu}{d_\mu\sum_{\nu\vdash N} m^2_\nu}} P_\mu.
\ee
This proves the statement of Theorem~\ref{optmeas}.

\section{Solution of primal and dual problem for deterministic scheme}
\label{AppB}
{\bf Solution of the primal problem}
Since our goal is to find optimal set of effects $\{\Pi_{\mathbf{i}}\}_{\mathbf{i}\in \mathcal{I}}$, we expect from them to have the same type of symmetries implied by the MPBT operator $\rho$ from~\eqref{eigenrho} and signals $\{\sigma_{\mathbf{i}}\}_{\mathbf{i}\in \mathcal{I}}$ given in~\eqref{sigma} (see also  papers~\cite{majenz,2020arXiv200811194L,StuNJP}). Having this in mind we assume that optimal measurements are the elements of the algebra $\mathcal{A}^{k}_d(n)$ studied in~\cite{stud2020A} and briefly highlighted in Section~\ref{connA}. We know from~\cite{Kopszak2020}, that in non-optimal case of the multi-port based teleportation protocol, Alice's measurements are of the form of square-root measurements:
\be
\forall \ \mathbf{i}\in \mathcal{I} \qquad \Pi_{\mathbf{i}}=\frac{1}{\sqrt{\rho}}\sigma_{\mathbf{i}}\frac{1}{\sqrt{\rho}}.
\ee
The support of the square-root measurements is always restricted to the supports of the signals $\sigma_{\mathbf{i}}$, making the main technical obstacle for effective computations and getting closed expressions for fidelity. Namely, there is no easy way of calculating inversion of the operator $\rho$ analytically. However, exploiting symmetries of the protocol we can block-diagonalise $\rho$ and compute its any function on every irreducible subspace separately. 
In our case we allow for more general form of sandwiching of the operators $\sigma_{\mathbf{i}}$. Namely, let us assume the following form of the optimal POVMs:
\be
\label{rot}
\forall \ \mathbf{i}\in \mathcal{I} \qquad \Pi_{\mathbf{i}}=\Pi \sigma_{\mathbf{i}} \Pi,
\ee
with
\be
\label{pi1}
\Pi=\sum_{\alpha \vdash N-k} 
\sum_{\mu\in\alpha}p_{\mu}(\alpha)F_{\mu}(\alpha),\quad p_{\mu}(\alpha)\geq 0,
\ee
where $F_{\mu}(\alpha)$ are eigen-projectors of MPBT operator given in expression~\eqref{eigenrho}. The similar reasoning allows us to restrict considered matrix $X_{A}$ from \eqref{ccoo} to the following form
\be
\label{XA}
X_{A}=O_A^{\dagger}O_A=\sum_{\mu \vdash N}c_{\mu}P_{\mu},\quad c_{\mu}\geq 0,
\ee
where $P_{\mu}$ are Young projectors for irreps labelled by $\mu \vdash N$.
Using assumed forms of the operators $\{\Pi_{\mathbf{i}}\}_{\mathbf{i}\in \mathcal{I}}$  and $X_A$, expression for $F^*$ in~\eqref{primF} reads as
\be
\label{to}
\begin{split}
	F^*=\frac{1}{d^{2k}}\max_{\{\Pi_{\mathbf{i}}\}}\sum_{\mathbf{i}\in\mathcal{I}}\tr\left(\Pi_{\mathbf{i}}\sigma_{\mathbf{i}} \right)=\frac{1}{d^{2k}}\max_{\Pi}\sum_{\mathbf{i}\in\mathcal{I}}\tr\left(\Pi \sigma_{\mathbf{i}} \Pi \sigma_{\mathbf{i}}\right)=\frac{k!\binom{N}{k}}{d^{2k}}\max_{\Pi}\tr\left(\Pi \sigma_{\mathbf{i}_0} \Pi \sigma_{\mathbf{i}_0}\right).
\end{split}
\ee
Substituting~\eqref{pi1} to~\eqref{to} and using canonical signal $\sigma_{\mathbf{i}_0}=(1/d^N)V^{(k)}$ from~\eqref{canonical}, we expand the above expression to
\be
\begin{split}
	F^*=\frac{k!\binom{N}{k}}{d^{2N+2k}}\max_{\substack{\{p_{\mu}(\alpha)\}\\ \{p_{\nu}(\beta)\}}}\sum_{\alpha,\beta \vdash N-k} 
	\sum_{\substack{\mu\in\alpha\\ \nu\in\beta}}p_{\mu}(\alpha)p_{\nu}(\beta)\tr\left(F_{\mu}(\alpha)V^{(k)}F_{\nu}(\beta) V^{(k)}\right).
\end{split}
\ee
Computing the trace $\tr\left(F_{\mu}(\alpha)V^{(k)}F_{\nu}(\beta) V^{(k)}\right)$ we have
\be
\label{Ff}
\begin{split}
	F^*&=\frac{k!\binom{N}{k}}{d^{2N+2k}}\max_{\substack{\{p_{\mu}(\alpha)\}\\ \{p_{\nu}(\beta_2)\}}}\sum_{\alpha,\beta \vdash N-k} 
	\sum_{\substack{\mu\in\alpha\\ \nu\in\beta}}p_{\mu}(\alpha)p_{\nu}(\beta)\tr\left[ P_{\alpha}P_{\mu}P_{\beta}\tr_{n-2,n-3}\left(P_{\nu} \right)V^{(k)}  \right]\\
	&=\frac{k!\binom{N}{k}}{d^{2N+2k}}\max_{\substack{\{p_{\mu}(\alpha)\}\\ \{p_{\nu}(\beta)\}}}\sum_{\alpha,\beta \vdash N-k} 
	\sum_{\substack{\mu\in\alpha\\ \nu\in\beta}}p_{\mu}(\alpha)p_{\nu}(\beta)m_{\nu/\beta}\frac{m_{\nu}}{m_{\beta}}\tr(\underbrace{P_{\beta}P_{\alpha}}_{\delta_{\beta \alpha}P_{\alpha}}P_{\mu})\\
	&=\frac{k!\binom{N}{k}}{d^{2N+2k}}\max_{\substack{\{p_{\mu}(\alpha)\}\\ \{p_{\nu}(\alpha)\}}}\sum_{\alpha\vdash N-k} 
	\sum_{\substack{\mu,\nu \in\alpha}}p_{\mu}(\alpha)p_{\nu}(\alpha)m_{\nu/\alpha}\frac{m_{\nu}}{m_{\alpha}}\tr\left(P_{\alpha}P_{\mu}\right)\\
	&=\frac{k!\binom{N}{k}}{d^{2N+2k}}\max_{\substack{\{p_{\mu}(\alpha)\}\\ \{p_{\nu}(\alpha)\}}}\sum_{\alpha\vdash N-k}\frac{d_{\alpha}}{m_{\alpha}} 
	\sum_{\substack{\mu,\nu\in\alpha}}p_{\mu}(\alpha)p_{\nu}(\alpha)m_{\mu/\alpha}m_{\nu/\alpha}m_{\nu}m_{\mu}\\
	&=\frac{k!\binom{N}{k}}{d^{2N+2k}}\max_{\substack{\{p_{\mu}(\alpha)\}}}\sum_{\alpha\vdash N-k}\frac{d_{\alpha}}{m_{\alpha}}\left(\sum_{\mu\in\alpha}p_{\mu}(\alpha)m_{\mu/\alpha}m_{\mu} \right)^2.
\end{split}
\ee
The above calculations require a justification. Namely, the expression under the trace in the first line comes by exploiting Lemma 21 and Fact 1 in~\cite{stud2020A}, while the expressions in the second and fourth line hold because of Corollary 10 and discussion below it in the same paper.
Now, using fact that the MPBT operator is a linear combination of all signals $\sigma_{\mathbf{i}}, \ \mathbf{i}\in \mathcal{I}$, together with spectral decomposition of $\rho$ from~\eqref{eigenrho}, we observe 
\be
\label{130}
\sum_{\mathbf{i}\in\mathcal{I}}\Pi_{\mathbf{i}}=\sum_{\mathbf{i}\in\mathcal{I}}\Pi \sigma_{\mathbf{i}} \Pi=\Pi \sum_{\mathbf{i}\in\mathcal{I}}\sigma_{\mathbf{i}} \Pi=\Pi\rho\Pi=\Pi^2\rho=\sum_{\alpha \vdash N-k} 
\sum_{\mu\in\alpha}p^2_{\mu}(\alpha)\lambda_{\mu}(\alpha)F_{\mu}(\alpha).
\ee
Having explicit form~\eqref{130}, we substitute it to the first constraint in~\eqref{ccoo}, getting
\be
\sum_{\alpha \vdash N-k} 
\sum_{\mu\in\alpha}p^2_{\mu}(\alpha)\lambda_{\mu}(\alpha)F_{\mu}(\alpha)=\sum_{\mu}\sum_{\alpha\in \mu}p^2_{\mu}(\alpha)\lambda_{\mu}(\alpha)F_{\mu}(\alpha)\leq \sum_{\mu \vdash N}c_{\mu}P_{\mu}\ot \mathbf{1}_{B}.
\ee
The right hand side is obtained by using assumed form of $X_{A}$ in~\eqref{XA} to second constraint in~\eqref{ccoo}. Observing that $F_{\mu}(\alpha)\subset P_{\mu}$, we deduce $p_{\mu}^2(\alpha)\lambda_{\mu}(\alpha)\leq c_{\mu}$. Moreover, $F^*$ can increase only when the coefficients $p_{\mu}(\alpha)$ increase, so for fixed $c_{\mu}$ we get the maximal possible value when
\be
\label{max0}
\forall \ \alpha  \qquad p_{\mu}^2(\alpha)\lambda_{\mu}(\alpha)= c_{\mu}.
\ee
From the second constraint in~\eqref{ccoo}, together with assumed form of $X_{\mathbf{A}}$ in~\eqref{XA} we write
\be
\tr X_{A}=\sum_{\mu}c_{\mu}\tr P_{\mu}=\sum_{\mu}c_{\mu}d_{\mu}m_{\mu}=d^N.
\ee
Defining a new auxiliary variable $v_{\mu}^2\equiv \frac{1}{d^N}c_{\mu}m_{\mu}d_{\mu}$, we rewrite~\eqref{max0} as
\be
\frac{1}{d^N}p_{\mu}^2(\alpha)\lambda_{\mu}(\alpha)m_{\mu}d_{\mu}=\frac{1}{d^N}c_{\mu}m_{\mu}d_{\mu}=v_{\mu}^2.
\ee
Using explicit expression for the eigenvalues $\lambda_{\mu}(\alpha)$ from~\eqref{eig} we get
\be
\label{pp}
p_{\mu}(\alpha)=\frac{d^N}{\sqrt{k!\binom{N}{k}}}\sqrt{\frac{m_{\alpha}}{d_{\alpha}}}\frac{v_{\mu}}{m_{\mu}}.
\ee
Having the above we rewrite~\eqref{Ff} as
\be
\label{FFf}
\begin{split}
	F^*&=\frac{k!\binom{N}{k}}{d^{2N+2k}}\max_{\substack{\{p_{\mu}(\alpha)\}}}\sum_{\alpha\vdash N-k}\frac{d_{\alpha}}{m_{\alpha}}\left(\sum_{\mu\in\alpha}p_{\mu}(\alpha)m_{\mu/\alpha}m_{\mu} \right)^2=\frac{1}{d^{2k}}\max_{\{v_{\mu}\}}\sum_{\alpha}\left(\sum_{\mu\in\alpha}m_{\mu/\alpha}v_{\mu} \right)^2,
\end{split}
\ee
where the symbol  $m_{\mu/\alpha}$ denotes in how many ways we can get Young frame $\mu$ from Young frame $\alpha$ by adding $k$ boxes in a proper way one by one. Let us rewrite sums over irreps as follows
\be
\label{expand}
\begin{split}
\sum_{\alpha}\left(\sum_{\mu\in\alpha}m_{\mu/\alpha}v_{\mu} \right)^2&=\left(\sum_{\alpha}\sum_{\mu\in\alpha}m^2_{\mu/\alpha}v^2_{\mu}+\sum_{\alpha}\sum_{\substack{\mu,\nu\in\alpha \\ \mu\neq \nu}}m_{\mu/\alpha}m_{\nu/\alpha}v_{\mu}v_{\nu}\right)\\
&=\sum_{\mu}\left(\sum_{\alpha \in \mu}m^2_{\mu/\alpha}\right)v_{\mu}^2+\sum_{\substack{\mu\neq \nu \\\mu \sim_k \nu}}\left(\sum_{\alpha \in \mu \wedge \nu}m_{\mu/\alpha}m_{\nu/\alpha}\right)v_{\mu}v_{\nu}\\
&=\sum_{\mu}M_{\mu\mu}v^2_{\mu}+\sum_{\substack{\mu\neq \nu \\\mu \sim_k \nu}}M_{\mu\nu}v_{\mu}v_{\nu}.
\end{split}
\ee
The symbol $\mu \sim_k \nu$ means Young frame $\nu$ can be obtained from Young frame $\mu$ by moving up to $k$ boxes, while the symbol $\alpha \in \mu \wedge \nu$ means Young frame $\alpha$ can be obtained from Young frame $\mu$ as well as from $\nu$ by subtracting $k$ boxes. 
Observing that $\sum_{\mu}v_{\mu}^2=1$, we have maximisation over all vectors $v=(v_{\mu})$, such that $||v||=1$ in~\eqref{FFf}. This, together with~\eqref{expand} allows us to write
\be
F^*=\frac{1}{d^{2k}}\max_{v: ||v||=1}\left(\sum_{\mu}M_{\mu\mu}v^2_{\mu}+\sum_{\substack{\mu\neq \nu \\\mu \underset{k}{\sim} \nu}}M_{\mu\nu}v_{\mu}v_{\nu}\right)=\frac{1}{d^{2k}}\max_{v: ||v||=1}\<v|M_F^{d,k}|v\>=\frac{1}{d^{2k}} \lambda_{\max}(M_F^{d,k}),
\ee
where $M_F^{d,k}$ is the generalised teleportation matrix given through Definition~\ref{gen_MF}. This finishes the proof for the primal SDP problem.\\
{\bf Solution of the dual problem} Since here we consider the dual problem, it is natural to assume that it has the same type of symmetries as the primal one. In particular, we assume that the operator $\Omega$ from~\eqref{Fdual} is an element of algebra $\mathcal{A}^{(k)}_n(d)$. In particular, we assume it is a linear combination of projectors $F_{\mu}(\alpha)$, which are the eigen-projectors for MPBT operator, see~\eqref{eigenrho}, with some unknown coefficients which we have to optimise.  First, for further, purely technical purposes let us introduce an auxiliary operator $\widetilde{\Omega}$  which is an element of the algebra $\mathcal{A}^{(k)}_n(d)$:
\be
\widetilde{\Omega}=\sum_{\alpha \vdash N-k}\widetilde{\Omega}(\alpha)=\sum_{\alpha \vdash N-k}\sum_{\mu\in\alpha}\omega_{\mu}(\alpha)F_{\mu}(\alpha),\qquad \omega_{\mu}(\alpha)>0,
\ee
where $\omega_{\mu}(\alpha)$ are the coefficients which we have to determine to get the minimal value of $F_{*}$ in~\eqref{Fdual}, and satisfy constraint~\eqref{cons_dual}. Let us define coefficients
\be
\forall \ \alpha \vdash N-k \quad c(\alpha)\equiv\frac{1}{d^N}\sum_{k=1}^{d_{\alpha}}\sum_{l=1}^{m_{\alpha}}\tr\left[|\varphi_{k,l}(\alpha)\>\<\varphi_{k,l}(\alpha)|\widetilde{\Omega}^{-1}(\alpha)V^{(k)}\right],
\ee
where vectors $|\varphi_{k,l}(\alpha)\>$ for fixed $l$ span orthonormal basis in irrep $\alpha$ of $S(N-k)$ with the dimension  $d_{\alpha}$ and multiplicity $m_{\alpha}$ in the Schur-Weyl duality. Now we apply Theorem 1 and Lemma 1 from paper~\cite{Lewy} getting a relation
\be
\label{c1}
\forall \ \alpha \vdash N-k \quad \sum_{\mu\in\alpha}\omega_{\mu}(\alpha)F_{\mu}(\alpha)-\frac{1}{c(\alpha)}P_{\alpha}V^{(k)}\geq 0.
\ee
The term $P_{\alpha}V^{(k)}$ is just projection of the unnormalised canonical signal state $\sigma_{\mathbf{i}_0}$ from~\eqref{canonical} on irrep $\alpha$ by Young projector $P_{\alpha}$. It is enough to consider only $\sigma_{\mathbf{i}_0}$ since all the signals $\{\sigma_{\mathbf{i}}\}_{\mathbf{i}\in\mathcal{I}}$, as well as the operator $\widetilde{\Omega}$ are invariant under elements from the coset $\mathcal{S}_{n,k}\equiv \frac{S(n-k)}{S(n-2k)}$ and transformations $U^{\ot (n-k)}\ot \overline{U}^{\ot k}$. Evaluation of the coefficients $c(\alpha)$ is presented in Lemma~\ref{coef}, the explicit expression for them is of the form
\be
\forall \ \alpha\vdash N-k \quad \forall \ \mu\in\alpha \qquad c(\alpha)=\frac{1}{d^N}\sum_{\mu\in\alpha}\omega^{-1}_{\mu}(\alpha)m_{\mu/\alpha}\frac{m_{\mu}}{m_{\alpha}}\geq 0,
\ee
where $m_{\mu/\alpha}$ denotes number of all possibilities of adding $k$ boxes in a proper way to $\alpha$ in order to get $\mu$. Multiplying both sides of~\eqref{c1} by $c(\alpha)$ we have
\be
c(\alpha)\sum_{\mu\in\alpha}\omega_{\mu}(\alpha)F_{\mu}(\alpha)-P_{\alpha}V^{(k)}\geq 0.
\ee
Defining the operator $\Omega$ in the minimisation problem~\eqref{Fdual} as
\be
\Omega\equiv\sum_{\alpha}\sum_{\mu\in\alpha}c(\alpha)\omega_{\mu}(\alpha)F_{\mu}(\alpha)
\ee
we automatically satisfy the constraint~\eqref{cons_dual}, having for every $\alpha$ relation:
\be
\sum_{\mu\in\alpha}c(\alpha)\omega_{\mu}(\alpha)F_{\mu}(\alpha)-P_{\alpha}V^{(k)}\geq 0.
\ee
Having explicit form  of the allowed operator $\Omega$ we plug it in expression~\eqref{Fdual}:
\be
F_{*}=d^{N-2k}\min\limits_{\Omega}||\tr_{(k)}\Omega||_{\infty}=d^{N-2k}\operatorname{min}||\sum_{\alpha \vdash N-k}\sum_{\mu\in\alpha}c(\alpha)\omega_{\mu}(\alpha)\tr_{(k)}F_{\mu}(\alpha)||_{\infty}.
\ee
The symbol $\tr_{(k)}$ denotes partial trace over last $k$ particles. Applying Lemma 19 from~\cite{stud2020A}, saying that $\tr_{(k)}F_{\mu}(\alpha)=m_{\mu/\alpha}\frac{m_{\alpha}}{m_{\mu}}P_{\mu}$, and taking into account that $P_{\mu}$ are projectors, we write $F_{*}$ as
\be
\label{above}
\begin{split}
F_{*}&=d^{N-2k}\min \max_{\mu}\sum_{\alpha \in \mu}c(\alpha)\omega_{\mu}(\alpha)m_{\mu/\alpha}\frac{m_{\alpha}}{m_{\mu}}=\frac{1}{d^{2k}}\min\max_{\mu}\sum_{\alpha\in\mu}\sum_{\nu\in\alpha} m_{\mu/\alpha}m_{\nu/\alpha}\frac{m_{\nu}}{m_{\mu}}\frac{\omega_{\mu}(\alpha)}{\omega_{\nu}(\alpha)}.
\end{split}
\ee
Introducing a new variable $\forall \alpha \ \forall \mu\in\alpha \quad t_{\mu}(\alpha):=m_{\mu}/\omega_{\mu}(\alpha)$, we rewrite~\eqref{above} in the following form:
\be
F_{*}=\frac{1}{d^{2k}}\min\max_{\mu}\sum_{\alpha \in\mu}\sum_{\nu\in\alpha} m_{\mu/\alpha}m_{\nu/\alpha}\frac{t_{\nu}(\alpha)}{t_{\mu}(\alpha)}=\frac{1}{d^{2k}}\min\max_{\mu}\sum_{\alpha \in\mu}m_{\mu/\alpha}\frac{\sum_{\nu\in\alpha}m_{\nu/\alpha}t_{\nu}(\alpha)}{t_{\mu}(\alpha)}.
\ee
Since we are looking for any feasible solution we can assume $\forall \ \alpha \ t_{\mu}(\alpha)=t_{\mu}$:
\be
F_{*}=\frac{1}{d^{2k}}\min\limits_{\{t_{\mu}\}}\max_{\mu}\sum_{\alpha \in\mu}m_{\mu/\alpha}\frac{\sum_{\nu\in\alpha}m_{\nu/\alpha}t_{\nu}}{t_{\mu}}=\frac{1}{d^{2k}}\min\limits_{\{t_{\mu}\}}\max_{\mu}\sum_{\alpha \in \mu}m_{\mu/\alpha}\frac{\sum_{\nu\in\alpha}m_{\nu/\alpha}t_{\nu}}{t_{\mu}}.
\ee
Now, recalling Definition~\ref{gen_MF} of the generalised teleportation matrix $M_F^{d,k}$ we observe that
\be
\min\limits_{\{t_{\mu}\}}\max\limits_{\mu} \sum_{\alpha \in \mu}m_{\mu/\alpha}\frac{\sum_{\nu\in\alpha}m_{\nu/\alpha}t_{\nu}}{t_{\mu}}=\min\limits_{\{t_{\mu}\}}\max\limits_{\mu}\frac{\sum_{\nu}(M_F^{d,k})_{\mu\nu}t_{\nu}}{t_{\mu}}=\lambda_{\max}(M_F^{d,k}).
\ee
\vspace{0.3cm}\\
Finally, let us notice that the solutions of primal and dual problems allow us to deduce the explicit form of the optimal measurements performed by Alice, as well as explicit form of the rotation $O_A$ from~\eqref{resource0}. Namely, expression~\eqref{pp} together with~\eqref{rot} and~\eqref{pi1} gives us 
\be
\Pi=\frac{d^N}{\sqrt{k!\binom{N}{k}}}\sum_{\alpha \vdash N-k} 
\sum_{\mu\in\alpha}\sqrt{\frac{m_{\alpha}}{d_{\alpha}}}\frac{v_{\mu}}{m_{\mu}}F_{\mu}(\alpha),
\ee
while expressions~\eqref{pp} and~\eqref{max0}, together with~\eqref{eig} imply
\be
O_A= \sqrt{d^N}\sum_\mu\frac{v_\mu}{d_\mu m_\mu} P_\mu.
\ee
The vector $v=(v_{\mu})$ is an eigenvector of the generalised teleportation matrix, corresponding to its maximal eigenvalue. This numbers can be computed effectively due to analysis presented in~\cite{StuNJP}. This proves Theorem~\ref{optmeasurements} from Section~\ref{mainPBT}.

\section{Proof of Theorem~\ref{kN}}
\label{DkN}
The proof  is based on the statement of Theorem~\ref{Thmp}. Namely, let us notice that expression~\eqref{p} can be rewritten as it is done below:
\begin{align}
\label{above2}
        p&=\frac{N!}{(d^2+N-1)!}\frac{(d^2+N-1-k(N))!}{(N-k(N))!}\\
&=\frac{1}{(d^2+N-1)\cdots(N+1)}\frac{(d^2+N-1-k(N))!}{(N-k(N))!}\\
&=\frac{(d^2+N-1-k(N))\cdots(N+1-k(N))}{(d^2+N-1)\cdots(N+1)}\\
&=\prod_{i=2}^{d^2}\frac{N-1+i-k(N)}{N-1+i}=\prod_{i=2}^{d^2}\left(1-\frac{k(N)}{N-1+i}\right)
\end{align}
Now, if $k(N)=o(N)$, which means $k(N)/N\rightarrow 0$ for $N\rightarrow \infty$, the limit of expression~\eqref{above2} is
\begin{equation}
\lim_{N\rightarrow\infty}\prod_{i=2}^{d^2}\left(1-\frac{k(N)}{N-1+i}\right) = 1.
\end{equation}
This finishes the proof.

\section{Proof of Theorem~\ref{thm}}
\label{PT7}
To prove the statement of theorem, for the reader's convenience, we divide the whole proof into a few smaller propositions and lemmas and prove them separately as a separate building blocks.
We start from the following proposition:
\begin{proposition}
\label{prop1}
Let $m_{\nu}$ denotes multiplicities of irreps $\psi^{\nu}$ of $S(N)$ in decomposition~\eqref{decomp} of the swap representation~\eqref{swap}. Then the following relation holds:
	\be
	\sum_{\nu :h(\nu )\leq d}m_{\nu }^{2}=\frac{1}{N!}\sum_{\sigma \in
		S(N)}d^{2l_{N}(\sigma )}, 
	\ee
	where $l_{N}(\sigma )$ is the number of cycles in the permutation $\sigma $
	as a permutation of $S(N).$
\end{proposition}

\begin{proof}
From group theory of characters~\cite{Fulton1991-book-rep} we know that the character of the
representation $V_{N}^{d}:S(N)\rightarrow \operatorname{Hom}[(\mathbb{C}^{d})^{\otimes N}]$ is of the form
\begin{equation}
\label{char}
\chi ^{V_{N}^{d}}=\sum_{\nu :h(\nu )\leq d}m_{\nu }\chi ^{\nu },
\end{equation}
where $\chi ^{\nu }$ are characters of irreps of $S(N)$, and the sum runs over all Young diagrams $\nu$ of height $h(\nu)$ at most $d$. The scalar product of characters~\eqref{char} equals to
\begin{equation}
\label{a}
(\chi ^{V_{N}^{d}},\chi ^{V_{N}^{d}})\equiv \frac{1}{N!}\sum_{\sigma \in
S(N)}\chi ^{V_{N}^{d}}(\sigma )\chi ^{V_{N}^{d}}(\sigma ^{-1})=\sum_{\nu
:h(\nu )\leq d}(m_{\nu })^{2}.
\end{equation}%
On the other hand, denoting by $l_N(\sigma)$ the number of disjoint cycles in permutation $\sigma \in S(N)$,  we have 
\begin{equation}
\label{b}
\chi ^{V_{N}^{d}}(\sigma )=d^{l_{N}(\sigma )}\Rightarrow (\chi
^{V_{N}^{d}},\chi ^{V_{N}^{d}})=\frac{1}{N!}\sum_{\sigma \in
S(N)}d^{2l_{N}(\sigma )}.
\end{equation}
Comparing right-hand sides of~\eqref{a} and~\eqref{b} we get the statement of the proposition.
\end{proof}

Now, we prove result regarding function $l_N:S(N)\rightarrow \mathbb{N}$ returning number of disjoint cycles of a permutation  $\sigma \in S(N)$.

\begin{lemma}
\label{lemma2}
	Let $\sigma \in S(N)$ and  $\sigma =(aN)\pi :\pi \in S(N-1)$ for $a=1,\ldots,N,$ then the number of cycles $l_N(\sigma)$ in permutation $\sigma$ satisfies:
	\be
	l_{N}(\sigma )=\begin{cases}
		l_{N-1}(\pi ):a\neq N, \\ 
		l_{N-1}(\pi )+1:a=N.
	\end{cases}
	\ee
	For the non-trivial cosets $S(N)/S(N-1)$ the number of cycles is the same.
\end{lemma}

\begin{proof}
The statement of the lemma follows directly from the decomposition
of a permutation $\pi \in S(N-1)$ into disjoint cycles%
\begin{equation}
\label{cycles}
\pi =(a\ldots)(b\ldots)\cdots(s\ldots).
\end{equation}%
Computing composition of $\pi$ from~\eqref{cycles} with transposition $(aN)$ we get
\begin{equation}
(aN)\pi =(Na\ldots)(b\ldots)\cdots(s\ldots).
\end{equation}%
We see, the number of cycles is the same in both groups $S(N)$ and $S(N-1).$ If $a=N$, then obviously $\sigma =\pi$, but in this case the function $l_{N}(\sigma)$ returns one more cycle $(N)$ of length one.
\end{proof}
Having the above considerations one can derive:
\begin{proposition}
\label{1iteracja}
	Let us consider decomposition into irreps two swap representations $V_{N}^{d}$ and $V_{N-1}^{d}$
	\be
	V_{N}^{d}\cong \bigoplus _{\nu :h(\nu )\leq d}m_{\nu }\psi ^{\nu },\qquad
	V_{N-1}^{d}\cong \bigoplus _{\alpha :h(\alpha )\leq d}m_{\alpha }\varphi
	^{\alpha }, 
	\ee
	where $h(\cdot)$ denotes height of a Young diagram, and $m_{\alpha},m_{\nu}$ are  multiplicities of respective irreps in the Schur-Weyl duality. With this notation the following relation holds
	\be
	\label{relation}
	\frac{\sum_{\nu :h(\nu )\leq d}m_{\nu }^{2}}{\sum_{\alpha :h(\alpha )\leq
			d}m_{\alpha }^{2}}=\frac{d^{2}+N-1}{N}. 
	\ee
\end{proposition}

\begin{proof}
From Proposition~\ref{prop1} we have 
\begin{equation}
\sum_{\nu :h(\nu )\leq d}m_{\nu }^{2}=\frac{1}{N!}\sum_{\sigma \in
S(N)}d^{2l_{N}(\sigma )}=\frac{1}{N!}\sum_{a=1}^{N}\sum_{\pi \in
S(N-1)}d^{2l_{N}[(aN)\pi ]}.
\end{equation}%
Now, applying Lemma~\ref{lemma2} we get 
\begin{equation}
\sum_{\nu :h(\nu )\leq d}m_{\nu }^{2}=\frac{1}{N!}\sum_{\pi \in
S(N-1)}\left(\sum_{a=1}^{N-1}d^{2l_{N-1}(\pi )}+d^{2l_{N-1}(\pi )+2}\right)=\frac{%
d^{2}+N-1}{N}\frac{1}{(N-1)!}\sum_{\pi \in S(N-1)}d^{2l_{N-1}(\pi )}.
\end{equation}%
Again, applying Proposition~\ref{prop1} we have
\begin{equation}
\label{last}
\sum_{\nu :h(\nu )\leq d}m_{\nu }^{2}=\frac{d^{2}+N-1}{N}\sum_{\alpha
:h(\alpha )\leq d}m_{\alpha }^{2}.
\end{equation}
This finishes the proof.
\end{proof}
Now by a multiplication side by side equation~\eqref{last} from Proposition~\ref{1iteracja} by $N,N-1,\ldots,N-k$, where $1\leq k\leq N-1$, we get the statement of Theorem~\ref{Thmp}.

\section{Proof of Lemma~\ref{coef}}
\label{PL8}
First let us notice that by Lemma 21 in~\cite{stud2020A} we have $V^{(k)}F_{\mu}(\alpha)=V^{(k)}P_{\alpha}P_{\mu}$, so
\be
\tr\left[|\varphi_{k,l}(\alpha)\>\<\varphi_{r,s}(\alpha)|V^{(k)}F_{\mu}(\alpha)\right]=\tr\left[|\varphi_{k,l}(\alpha)\>\<\varphi_{r,s}(\alpha)|V^{(k)}P_{\alpha}P_{\mu}\right]=\tr\left[|\varphi_{k,l}(\alpha)\>\<\varphi_{r,s}(\alpha)|\tr_{(k)}(V^{(k)})P_{\alpha}P_{\mu}\right].
\ee
The last equality follows from the fact that on last $k$ particles only the operator $V^{(k)}$ acts non-trivially. Computing the partial trace from $V^{(k)}$ and taking into account resolution of $P_{\alpha}=\sum_{p=1}^{d_{\alpha}}\sum_{q=1}^{m_{\alpha}}|\varphi_{p,q}(\alpha)\>\<\varphi_{p,q}(\alpha)|$, here $d_{\alpha}, m_{\alpha}$ are dimension and multiplicity of irrep $\alpha$ respectively, we have 
\be
\begin{split}
\tr\left[|\varphi_{k,l}(\alpha)\>\<\varphi_{r,s}(\alpha)|\tr_{(k)}(V^{(k)})P_{\alpha}P_{\mu}\right]&=\tr\left[|\varphi_{k,l}(\alpha)\>\<\varphi_{r,s}(\alpha)|\sum_{p=1}^{d_{\alpha}}\sum_{q=1}^{m_{\alpha}}|\varphi_{p,q}(\alpha)\>\<\varphi_{p,q}(\alpha)|P_{\mu}\right]\\
&=\sum_{p=1}^{d_{\alpha}}\sum_{q=1}^{m_{\alpha}}\delta_{rp}\delta_{sq}\tr\left[|\varphi_{k,l}(\alpha)\>\<\varphi_{p,q}(\alpha)|P_{\mu}\right]\\
&=\tr\left[|\varphi_{k,l}(\alpha)\>\<\varphi_{r,s}(\alpha)|P_{\mu}\right].
\end{split}
\ee
Let us notice that operator $P_{\mu}$ acts on $N$ particles, while $|\varphi_{k,l}(\alpha)\>\<\varphi_{r,s}(\alpha)|$ only on $N-k$ first systems. This property allows us to write $\tr\left[|\varphi_{k,l}(\alpha)\>\<\varphi_{r,s}(\alpha)|P_{\mu}\right]=\tr\left[|\varphi_{k,l}(\alpha)\>\<\varphi_{r,s}(\alpha)|\tr_{(k)}(P_{\mu})\right]$. Applying Corollary 10 from~\cite{stud2020A} saying that $\tr_{(k)}P_{\mu}=\sum_{\beta\in\mu}m_{\mu/\beta}\frac{m_{\mu}}{m_{\beta}}P_{\beta}$, we have
\be
\begin{split}
\tr\left[|\varphi_{k,l}(\alpha)\>\<\varphi_{r,s}(\alpha)|P_{\mu}\right]&=\sum_{\beta\in\mu}m_{\mu/\beta}\frac{m_{\mu}}{m_{\beta}}\tr\left[|\varphi_{k,l}(\alpha)\>\<\varphi_{r,s}(\alpha)|P_{\beta}\right]\\
&=\sum_{\beta\in\mu}\sum_{p=1}^{d_{\beta}}\sum_{q=1}^{m_{\beta}}m_{\mu/\beta}\frac{m_{\mu}}{m_{\beta}}\<\varphi_{r,s}(\alpha)|\varphi_{p,q}(\beta\>\<\varphi_{p,q}(\beta)|\varphi_{k,l}(\alpha)\>.
\end{split}
\ee
Using orthonormality of the basis vectors we obtain the statement.

\bibliographystyle{plainnat}
\bibliography{biblio2}

\end{document}